\documentclass[preprint]{aastex62}

\usepackage{xcolor}

\shorttitle{TIMES II: Turbulence and star formation activities}
\shortauthors{Yun et al.}

\begin{document}

\title{TIMES II: Investigating the Relation Between Turbulence and Star-forming Environments in Molecular Clouds}

\author[0000-0001-6842-1555]{Hyeong-Sik Yun}
\affiliation{School of Space Research, Kyung Hee University, 1732 Deogyeong-daero, Giheung-gu, Yongin-si, Gyeonggi-do, 17104, Republic of Korea\\(jeongeun.lee@khu.ac.kr, hs-yun@khu.ac.kr)}
\author[0000-0003-3119-2087]{Jeong-Eun Lee}
\affiliation{School of Space Research, Kyung Hee University, 1732 Deogyeong-daero, Giheung-gu, Yongin-si, Gyeonggi-do, 17104, Republic of Korea\\(jeongeun.lee@khu.ac.kr, hs-yun@khu.ac.kr)}
\author{Neal J. Evans II}
\affiliation{Department of Astronomy, The University of Texas at Austin, 2515 Speedway, Austin, TX 78712, USA}
\affiliation{Korea Astronomy and Space Science Institute, 776, Daedeok-daero, Yuseong-gu, Daejeon, 34055, Republic of Korea}
\author{Stella S. R. Offner}
\affiliation{Department of Astronomy, The University of Texas at Austin, 2515 Speedway, Austin, TX 78712, USA}
\author{Mark H. Heyer}
\affiliation{Department of Astronomy, University of Massachusetts Amherst, 710 N. Pleasant Street, Amherst, MA 01003, USA}
\author{Jungyeon Cho}
\affiliation{Department of Astronomy and Space Science, Chungnam National University, 99, Daehak-ro Yuseong-gu, Daejeon, 34134, Republic of Korea}
\author{Brandt A. L. Gaches}
\affiliation{Physikalisches Institut, Universit\"{a}t zu K\"{o}ln, Z\"{u}lpicher Stra$\beta$e 77, 50937, K\"{o}ln, Germany}
\affiliation{Center of Planetary Systems Habitability, The University of Texas at Austin, 2515 Speedway, Austin, TX 78712, USA}
\author{Yao-Lun Yang}
\affiliation{Department of Astronomy, University of Virginia, Charlottesville, VA 22904-4235, USA} 
\affiliation{RIKEN Cluster for Pioneering Research, Wako-shi, Saitama, 351-0106, Japan}
\author{How-Huan Chen}
\affiliation{Department of Astronomy, The University of Texas at Austin, 2515 Speedway, Austin, TX 78712, USA}
\author{Yunhee Choi}
\affiliation{Korea Astronomy and Space Science Institute, 776, Daedeok-daero, Yuseong-gu, Daejeon, 34055, Republic of Korea}
\author{Yong-Hee Lee}
\affiliation{School of Space Research, Kyung Hee University, 1732 Deogyeong-daero, Giheung-gu, Yongin-si, Gyeonggi-do, 17104, Republic of Korea\\(jeongeun.lee@khu.ac.kr, hs-yun@khu.ac.kr)}
\author{Giseon Baek}
\affiliation{School of Space Research, Kyung Hee University, 1732 Deogyeong-daero, Giheung-gu, Yongin-si, Gyeonggi-do, 17104, Republic of Korea\\(jeongeun.lee@khu.ac.kr, hs-yun@khu.ac.kr)}
\author{Minho Choi}
\affiliation{Korea Astronomy and Space Science Institute, 776, Daedeok-daero, Yuseong-gu, Daejeon, 34055, Republic of Korea}
\author{Jongsoo Kim}
\affiliation{Korea Astronomy and Space Science Institute, 776, Daedeok-daero, Yuseong-gu, Daejeon, 34055, Republic of Korea}
\author{Hyunwoo Kang}
\affiliation{Korea Astronomy and Space Science Institute, 776, Daedeok-daero, Yuseong-gu, Daejeon, 34055, Republic of Korea}
\author{Seokho Lee}
\affiliation{National Astronomical Observatory of Japan, 2-21-1 Osawa, Mitaka, Tokyo 181-8588, Japan}
\author{Ken'ichi Tatematsu}
\affiliation{Nobeyama Radio Observatory, National Astronomical Observatory of Japan, National Institutes of Natural Sciences, 462-2 Nobeyama, Minamimaki, Minamisaku, Nagano 384-1305, Japan}
\affiliation{Department of Astronomical Science, SOKENDAI (The Graduate University for Advanced Studies), 2-21-1 Osawa, Mitaka, Tokyo 181-8588, Japan}

\begin{abstract}
We investigate the effect of star formation on turbulence in the Orion A and Ophiuchus clouds using principal component analysis (PCA). We measure the properties of turbulence by applying PCA on the spectral maps in $^{13}$CO, C$^{18}$O, HCO$^+$ $J=$1$-$0, and CS $J=$2$-$1. First, the scaling relations derived from PCA of the $^{13}$CO maps show that the velocity difference ($\delta v$) for a given spatial scale ($L$) is the highest in the integral shaped filament (ISF) and L1688, where the most active star formation occurs in the two clouds. The $\delta v$ increases with the number density and total bolometric luminosity of the protostars in the sub-regions. Second, in the ISF and L1688 regions, the $\delta v$ of C$^{18}$O, HCO$^+$, and CS are generally higher than that of $^{13}$CO, which implies that the dense gas is more turbulent than the diffuse gas in the star-forming regions; stars form in dense gas, and dynamical activities associated with star formation, such as jets and outflows, can provide energy into the surrounding gas to enhance turbulent motions.
\end{abstract}


\section{Introduction} \label{Sec_intro}
Turbulence is a ubiquitous phenomenon in the interstellar medium and plays a crucial role in the evolution of molecular clouds \citep{Mac03,Elm04}. A large-scale turbulent motion with a supersonic speed produces shocks that make density and velocity fluctuations in molecular clouds (MCs). Within these fluctuations, high-density regions evolve to clumps and cores via gravitational collapse \citep{Pad01b,Mac04,Kle04}. On the other hand, large-scale supersonic turbulence can also prevent global collapse and support the structure of MCs \citep{Mac04,Kle04,Fed15}. On small scales, the fragmentation of dense cores is also affected by turbulence \citep{Wal12}, and stars are actively formed after the dissipation of turbulence \citep{Mye83,Nak98,Goo98}.

Turbulence is known to be one of the essential factors determining the star formation rate (SFR) in the interstellar medium \citep{Fed12,Fed15,Fed18}. \citet{Fed15} showed that turbulence, as well as magnetic fields and stellar feedback should be included in numerical simulations to produce a realistic SFR. The SFR depends on how the turbulence energy is distributed between the compressive and solenoidal modes; the SFR is enhanced more by the compressive mode than the solenoidal mode \citep{Fed12,Fed18}. As a result, turbulence influences when and where stars form \citep{McK07,Hen12,Pad14}, and understanding the nature of turbulence is important to better understand the star formation in the interstellar medium.

After \citet{Lar81} found the power-law relation between linewidth and size of MCs, many studies have investigated the linewidth-size relations for MCs \citep{Sol87,Hey09} and cores \citep{Mye83,Goo98} using various molecular lines. The power-law indices of the linewidth-size relations have been compared with that of the velocity spectrum of turbulence to study a common feature of interstellar turbulent motion. \citet{Goo98} introduced four types of the linewidth-size relations (Type 1 to 4). These relations may have different slopes, and thus, provide information on different characteristics of interstellar turbulence. Among these types, the Type 4 relation obtained for a single-tracer for a single-cloud is probably the best diagnostic tool of the turbulence within a cloud \citep{Goo98}. 

Turbulence is statistically characterized by the way kinetic energy varies with a wavenumber ($k$) in Fourier space \citep[i.e., energy spectrum $E(k) \propto k^{-\beta}$,][]{Kol41}. For example, incompressible subsonic turbulence gives $\beta=5/3$ \citep{Kol41} while supersonic compressible turbulence gives $\beta=2$ \citep{Pas88,Gam96}. Unfortunately, obtaining an energy spectrum from molecular line maps is challenging \citep{Elm04}. One approach to obtain the properties of turbulence is measuring the velocity structure function using statistical methods \citep{Hey97,Kle00,Oss02,Elm04,Boy18,Xu20}. The 2nd order velocity structure function is related to the autocorrelation function, which in turn is the inverse Fourier transform of the power spectrum. Within the inertial range, the structure function follows a power-law relationship with spatial displacement, $l$, 
\begin{equation}
\sqrt{\langle v_l^{2} \rangle} \propto l^{\gamma}.
\end{equation}
A typically relation between the exponents $\gamma$ and $\beta$ is $\gamma = (\beta - 1)/2$. A recent study using a high-resolution simulation \citep{Fed21} showed that $\beta$ for subsonic and supersonic turbulence are 0.39$\pm$0.02 and 0.49$\pm$0.01, respectively, which are consistent with other theoretical studies \citep{Kol41,Pas88,Gam96}.

Principal component analysis \citep[PCA;][]{Hey97,Bru13} is one of the statistical methods to derive the underlying low order velocity structure function of turbulence from an observed spectral map \citep{Hey97,Bru03c,Bru02b,Hey04,Hey09,Rom11,Fed19}. PCA utilizes both spatial and velocity information and derives the power-law relation between a velocity difference ($\delta v$) and spatial scale ($L$) of turbulent motions, 
\begin{equation}
\delta v = \delta v_0 L^{\alpha}. \label{equ_scal}
\end{equation}
where $\delta v_0$ is $\delta v$ at 1~pc, and $\alpha$ is a PCA scaling exponent. The PCA scaling exponent is rescaled to 
$\gamma$ based on the calibration described by \citet{Bru02a} and \citet{Bru03c}. We refer to this power-law relation as a scaling relation of PCA. The scaling relation is considered to be one of the Type 4 relations because it describes variation of velocity difference traced by a single molecular line as a function of spatial scale within a cloud.

Using PCA, \citet{Hey04} derived the scaling relation from the spectral map of $^{12}$CO $J=$1$-$0 for each of 27 giant MCs. They found that the $\alpha$ and $\delta v_0$ values for 27 MCs are all similar each other ($\alpha$ and $\delta v_0$ are 0.62$\pm$0.09 and 0.90$\pm$0.9~km~s$^{-1}$, respectively). This mean value of $\alpha$ corresponds to a mean value of $\gamma$=0.51. The small scatter in $\alpha$ and $\delta v_0$ for this set of clouds strongly suggests that the velocity structure functions derived over the full extent of molecular clouds follow the same functional form, indicating that turbulence is universal in the molecular interstellar medium. However, \citet{Hey06} also found that the derived $\delta v_0$ values can vary with the local environment within a molecular cloud. This $\delta v_0$ difference could be related to the local energy dissipation or injection within the cloud such as stellar feedback from star-forming activities \citep{Boy16}. \citet{Koch17} carried out a systematic parameter study of molecular clouds using magneto-hydrodynamic simulations and found PCA is also sensitive to changes in the virial, plasma parameters, and the solenoidal fraction of the turbulence.

Other factors that can cause the $\delta v$ difference for a given $L$ is the uncertainty of distance \citep{Hey04}. PCA derives each $L$ via multiplying the angular size of turbulent motion by the distance to a cloud. If the distance is overestimated, PCA would overestimate $L$ and consequently underestimate $\delta v$ for a given $L$. Uncertain distances to MCs therefore cause the over and underestimation of $L$ resulting in a difference in $\delta v$ between the MCs. Therefore, knowing the accurate distance is important for investigating the turbulence scaling relation correctly. 

Many previous studies of turbulence have observed the $J$=1$-$0 transition of the $^{12}$CO and $^{13}$CO molecules \citep{Bal87,Hey92,Nag98,Shi11,Kon18}, which are the main tracers of molecular gas. But they can often fail to trace the entirety of the gas motions due to their optical thickness. Since cores and stars form in dense environments \citep{Pad01b,Mac04}, measuring turbulence in the dense parts of clouds is important to understand the relation between turbulence and star formation. Therefore, sampling MCs using multiple molecular transitions, which can trace different density environments, is necessary to investigate the properties of turbulence in the whole cloud \citep{Gac15}. This multi-transition study of a cloud can derive the Type 3 relation of \citet{Goo98}, which is associated with various density environments. By comparing the Type 3 relations between the MCs that have different star-forming environments, a relation between turbulence and star formation can be investigated \citep[the Type 1 relation of ][]{Goo98}. 

To assess the relation between turbulence and star formation, we carried out a systematic observation program for the Orion A and Ophiuchus clouds \citep[][Paper 1]{Yun21} using the 13.7-m telescope at the Taeduk Radio Astronomy Observatory \citep[TRAO; ][]{Jeo19,Roh99}. The TRAO telescope is an excellent facility to observe large areas in multiple molecular transitions efficiently. The Orion A and Ophiuchus clouds contain various star-forming environments: active massive and low-mass star formation occurs in the Orion A cloud \citep{Ike07,All08,Nak12,Meg12,Fur16}, and active low-mass star formation occurs in the Ophiuchus cloud \citep{Mot98,Wil08,Zha09,Dun15}. We observed these clouds in six molecular lines that trace different density environments. All the data were obtained via the TRAO Key Science Program (TRAO-KSP), ``mapping Turbulent properties In star-forming MolEcular clouds down to the Sonic scale" (TIMES; PI: Jeong-Eun Lee; Paper 1).

In this paper, we analyze the spectral maps obtained toward the Orion A and Ophiuchus clouds using PCA and investigate the scaling relations. We summarize the observed data in Section \ref{Sec_data}. Section \ref{Sec_pca} describes a methodology of PCA and the effect of a noise distribution on the PCA results. The results of PCA from the observed data are presented in Section \ref{Sec_rst}. In Section \ref{Sec_discs}, we will assess how the PCA results are related to the star formation activities and large-scale motion of the observed clouds. Section \ref{Sec_sum} summarizes the results of this paper. 

\section{Observations} \label{Sec_data}
We observed the Orion A and Ophiuchus clouds in six different molecular lines. The recent studies of distance using the \textit{Gaia} DR2 data reveal the detailed distance to the Orion A \citep{Gro18} and Ophiuchus clouds \citep{Zuc19}. The distance to the Orion A cloud varies from 391 to 467~pc along the filamentary structure \citep[see Table \ref{tbl_D_Ori}; ][]{Gro18}. Meanwhile, \citet{Zuc19} found a relatively constant distance across the Ophiuchus cloud, 144$\pm$7~pc on average. This distance value is consistent with the distances to L1688 and L1689, which were measured using VLBA parallax measurements \citep{Ort17}. We thus adopt the distance to L1688 from \citet{Ort17}, 137$\pm$1~pc, because of its small uncertainty. These detailed determination of distances make the Orion A and Ophiuchus clouds ideal targets to precisely assess the relation between turbulence and star formation environments.

The observation is carried out using the TRAO 13.7~m radio telescope as the TRAO-KSP, TIMES (Paper 1). The TRAO telescope has the SEQUOIA-TRAO receiver, which has 16 beams arranged in a 4$\times$4 array. The accessible frequency range is from 85 to 115~GHz with a fine spectral resolution of about 15~kHz resulting in a velocity resolution of 0.04~km~s$^{-1}$ at 110~GHz. The beam size is 46$\arcsec$ at 110~GHz. Also, SEQUOIA-TRAO allows access to two lines at 85$-$100~GHz or 100$-$115~GHz, simultaneously \citep{Jeo19}. 

All the spectral maps were obtained using the on-the-fly (OTF) mapping technique, which is efficient for observing a large area. We observed the MCs with not only $^{13}$CO $J$=1$-$0 but also C$^{18}$O $J$=1$-$0, HCN $J$=1$-$0, HCO$^+$ $J$=1$-$0, N$_2$H$^+$ $J$=1$-$0, and CS $J$=2$-$1, which trace gas in different density environments \citep{Gac15}. The pixel size and velocity resolution of the spectral maps are 20$\arcsec$ and 0.1~km~s$^{-1}$, respectively. All the spectral maps were obtained between 2016 and 2019. The details of the observation sequence and observed data were presented in Paper 1.

\section{Principal Component Analysis (PCA)} \label{Sec_pca}
To measure the properties of turbulence, we analyzed each of the observed spectral maps using principal component analysis \citep[PCA; ][]{Hey97,Bru13}. PCA is one of the multivariate analysis methods. Its formal role is describing multidimensional data as the linear combination of orthogonal principal components (PCs), where the PCs represent the most common features of the variation of the data. The eigenvectors generated by PCA provide a measure of velocity differences, $\delta v$, between observed line profiles in the map. For a given PC, the eigenimage, which is the dot product of each spectrum with the eigenvector, convey the spatial scale ($L$) over which the velocity difference occurs. The statistical error of this projection for a given position is equal to the rms of the spectrum.

The PCA produces $N_\mathrm{chan}$ orthogonal PCs where $N_\mathrm{chan}$ is the number of velocity channels in the cube data. From the PCs produced by the PCA, we have to consider only significant PCs in our further analysis. Figure \ref{fig_LSD_example} shows an example of the PCA for the $^{13}$CO line in the Orion A cloud; the ($\delta v$, $L$) points for the PCs up to the 32nd order are presented. Higher-order PCs represent increasingly smaller velocity differences that occur at increasingly smaller spatial scales in the cloud. If the eigenprojection values are indistinguishable from noise fluctuations, the measured velocity difference converge to a constant value (the red symbols in Figure \ref{fig_LSD_example}). Also, noisy patterns appear on their corresponding eigenimages (see Appendix \ref{App_scree}). Therefore, we adopt only the PCs (the black symbols) up to the order that starts to be insignificant or dominated by the noise. More discussion on how to select significant PCs are discussed in Appendix \ref{App_scree}. The percentage of variation ($p_\mathrm{var}$), which describes the fraction of the total variation covered by the adopted significant PCs, is defined by
\begin{equation}
p_\mathrm{var} = 100 \frac{\sum _\mathrm{i=1} ^{N_\mathrm{sig}} \lambda_\mathrm{i}}{\sum _\mathrm{i=1} ^{N_\mathrm{chan}} \lambda_\mathrm{i}}, \label{equ_pvar}
\end{equation}
where $N_\mathrm{sig}$ is the number of significant PCs and $\lambda_\mathrm{i}$ is an ith eigenvalue.

Among the significant PCs, the first-order (1st) PC reflects variance from spectral channels with signal against those spectral channels with only noise. Following \citet{Bru02b}, we thus omit the 1st PC. Also, the ($\delta v$, $L$) point of this component is often displaced from the best-fit power-law relation. The 2nd PC describes the largest velocity difference with the largest size. Therefore, this component often describes a large-scale systematic variation of velocity, such as the NW-SE velocity gradient in the Orion A cloud \citep{Hey92,Tat93,Shi11,Kon18}. In this case, we also exclude the ($\delta v$, $L$) point of the 2nd PC from the fit to avoid any probable contamination by the large-scale systematic motion.

The resulting ($\delta v$, $L$) points generally follow a power-law relation ($\delta v = \delta v_0L^{\alpha}$; a scaling relation). Refer to \citet{Hey97} and \citet{Bru13} for more details of the methodology of PCA. In this paper, the best-fit $\delta v_0$, $\alpha$, and their uncertainties for each of the scaling relations were estimated via the bootstrapping method with an orthogonal distance regression technique.

\begin{figure}
\epsscale{0.5}
\plotone{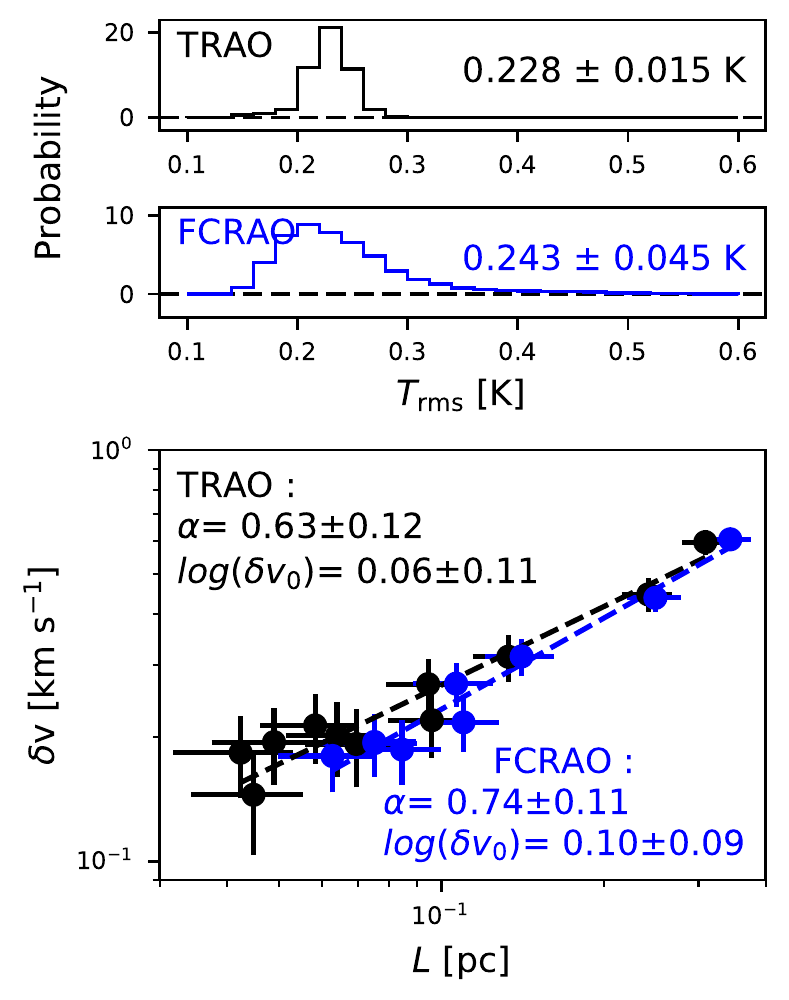}
\caption{The two probability distribution function (PDF) of the noise temperature ($T_\mathrm{rms}$) for the $^{13}$CO lines in the L1688 region, which were obtained using the TRAO telescope and the FCRAO telescope \citep{Rid06}, are presented in the top and middle panels, respectively. PCA scaling relations for the $^{13}$CO maps is shown in the bottom panel. The ${\delta}$v,L pairs derived from the eigenvectors and eigenimages for the TRAO and FCRAO data are presented in the black and blue circles, respectively. The error bars indicate the 1-$\sigma$ error. The power-law fitting result for each scaling relation is presented by the dashed line. The best-fit $\alpha$ and $\log(\delta v_0)$ values are summarized at the corners. \label{fig_rms_T_pdf_TnF}}
\end{figure}

PCA results are generally unaffected by variations of pixel size and velocity resolution \citep{Bru02a} but can be impacted by the level and inhomogeneity of the noise as differences in line profiles become indistinguishable from those generated by noise. We tested the effect of inhomogeneous noise on the PCA results by comparing our TRAO data set with prior observations having different noise properties. We adopted the $^{13}$CO data for the L1688 region in the Ophiuchus cloud obtained using the Five College Radio Astronomy Observatory (FCRAO) telescope as part of the Coordinated Molecular Probe Line Extinction and Thermal Emission Survey \citep[COMPLETE Survey;][]{Rid06}. Table \ref{tbl_data_params} shows the parameters of the TRAO and FCRAO data, such as the beam size, pixel size, velocity resolution, and rms noise temperature ($T_\mathrm{rms}$). The mean $T_\mathrm{rms}$ values for both data are similar. However, Figure \ref{fig_rms_T_pdf_TnF} shows that the probability distribution function (PDF) of $T_\mathrm{rms}$ for the FCRAO data is skewed to higher values while that of the TRAO data shows a Gaussian-like distribution. 

We assess the effect of the noise distribution on the PCA results by comparing the scaling relations for the TRAO and FCRAO data. The bottom panel of Figure \ref{fig_rms_T_pdf_TnF} shows the PCA scaling relations for the $^{13}$CO data obtained with the two radio antennae. The best-fit $\alpha$ and $\log(\delta v_0)$ values for the FCRAO data are consistent with those of the TRAO data within a 1-$\sigma$ range. Meanwhile, the eigenimage of the TRAO data exhibits the noisy pattern in a higher order PC (the 13th PC) compared to that of the FCRAO data (the 11th PC) resulting in different $N_\mathrm{sig}$: the PCA analysis found eleven PCs from the TRAO data while it found nine PCs from the FCRAO data. As a result, with these PCs, $p_\mathrm{var}$ for the TRAO data ($p_\mathrm{var}$=98.2~\%) is higher than that for the FCRAO data (97.4~\%). More significant PCs allow one to assess the $\delta v$ in smaller $L$. 

\section{Results} \label{Sec_rst}
\subsection{PCA Results for $^{13}$CO $J=$1$-$0} \label{Sec_regs}
\begin{figure}
\epsscale{0.5}
\plotone{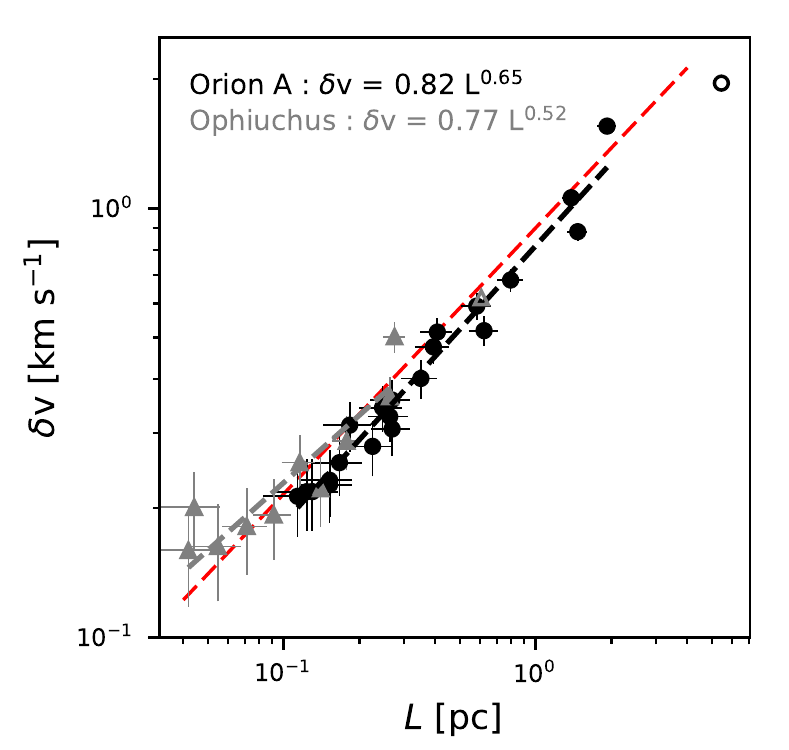}
\caption{The scaling relations for the $^{13}$CO line in the observed entire clouds. The black circles and gray triangles indicate PCs of the Orion A and Ophiuchus clouds, respectively. The open symbols represent the 2nd PCs that are not considered in the fitting process. The best-fit power-law relations are presented by the dashed lines and summarized on the upper-left corner. The scaling relation of the universality of turbulence \citep{Hey04} is also presented with the red dashed line. \label{fig_LSdiag_univ}}
\end{figure}

We applied the PCA method to the $^{13}$CO map of the Orion A and Ophiuchus clouds. A distance to the Orion A cloud is assumed as the average distance of 416.3~pc over the cloud \citep[][see Section 5.4 in Paper 1]{Kou18}. Figure \ref{fig_LSdiag_univ} shows the scaling relations for $^{13}$CO in the Orion A and Ophiuchus clouds. Table \ref{tbl_PCA_clouds} shows $N_\mathrm{sig}$, $p_\mathrm{var}$, and the best-fit $\delta v_0$ and $\alpha$ for the PCA of both clouds. The $\gamma$ values corresponding to the best-fit $\alpha$ values are 0.56 and 0.34 for the Orion A and Ophiuchus clouds, respectively. The set of $\delta v$, $L$ values lie close to the universal relationship (the red dashed line in Figure \ref{fig_LSdiag_univ}) derived by \citet{Hey04}.

The $p_\mathrm{var}$ for the Orion A cloud is lower than that for the Ophiuchus cloud (see Table \ref{tbl_PCA_clouds}). The $^{13}$CO map of the Orion A cloud has a wider velocity coverage (from -20 to 40~km~s$^{-1}$) than that for the Ophiuchus cloud (from -7 to 14~km~s$^{-1}$; Paper 1), including the velocity gradient along the Orion A filament. Therefore, a large portion of the cube data in the Orion A cloud is filled with channels dominated by noise, resulting in a greater contribution of noise to the total variation. The $p_\mathrm{var}$ value also varies with the total integrated intensity of the molecular line in the data. For example, the HCO$^+$ and CS lines in L1688 of the Ophiuchus cloud are weaker than those in the integral shaped filament (ISF) of the Orion A cloud (Paper 1), resulting in much lower $p_\mathrm{var}$ values (see Section \ref{PCA_lines}).

Adopting a single distance to the Orion A cloud would cause an over or underestimation of $\delta v$ at a given $L$; for example, if we adopt 416.3~pc, instead of 470~pc, for the distance to L1647-S \citep{Gro18}, $L$ would be underestimated resulting in an overestimation of $\delta v$ for a given $L$. Also, star formation environments vary depending on the position within the MC \citep{Ike07,All08,Nak12,Dun15}. Therefore, we divide the Orion A and Ophiuchus clouds into several sub-regions and investigated the PCA scaling relations of $^{13}$CO.

\begin{figure}
\epsscale{1.0}
\plotone{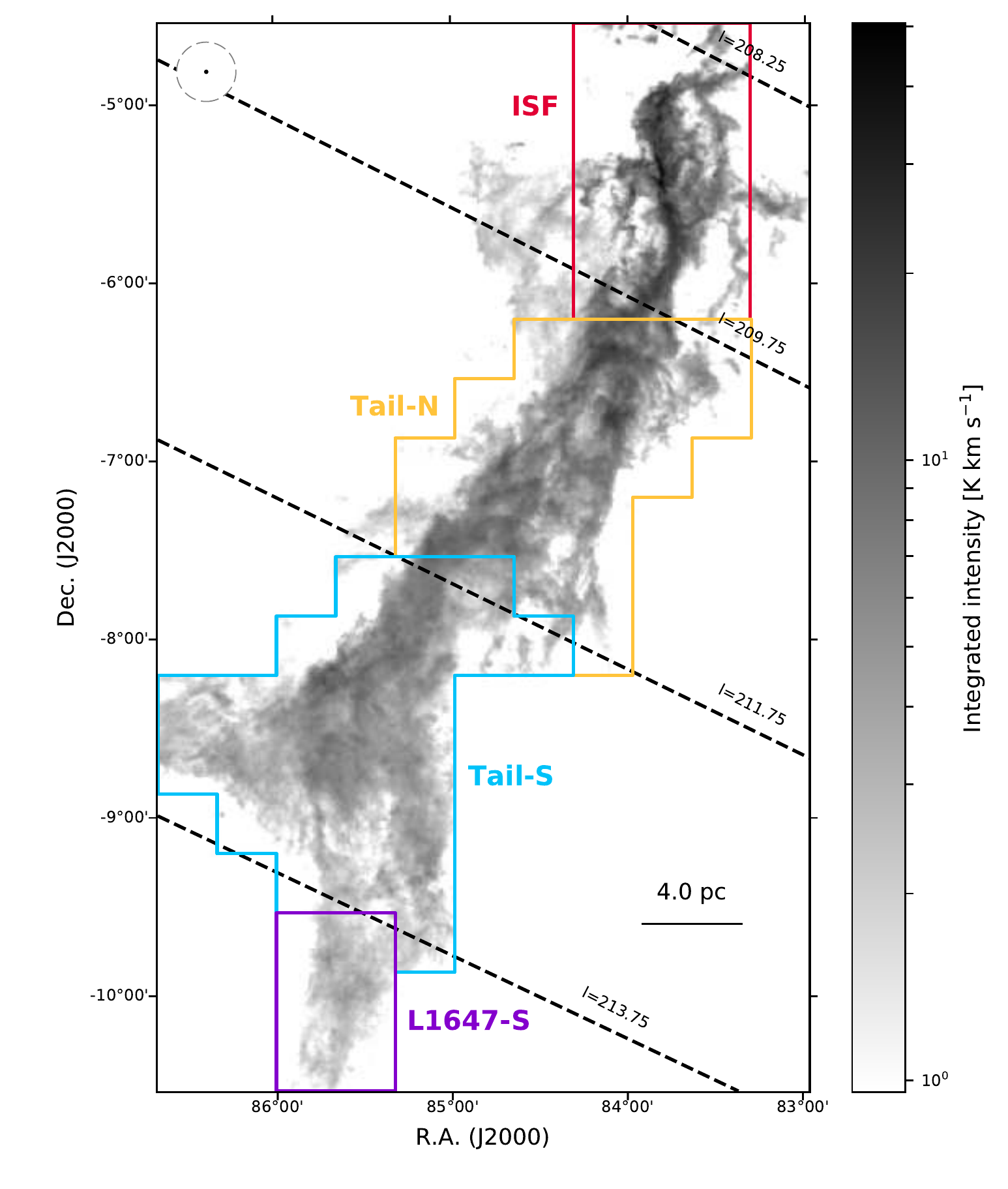}
\caption{Four sub-regions selected based on the distance in Table \ref{tbl_D_Ori}. Four selected sub-regions are outlined with the solid lines in different colors. The black dashed lines indicate the edges of the selected sub-regions in galactic longitude. Because the ranges of \textit{l} for the sub-regions are overlapped each other (0$\degr$.5; see Table \ref{tbl_D_Ori}), we adopted \textit{l} for the middle of the overlapped ranges (\textit{l}=208.25, 209.75, 211.75, and 213.75~$\degr$) as the edges of the sub-regions. The background image is the integrated intensity map of the $^{13}$CO line. The black dot in the gray dashed circle on the left-upper corner is the beam size. \label{fig_Ori13CO}}
\end{figure}

\begin{figure}
\epsscale{1.0}
\plotone{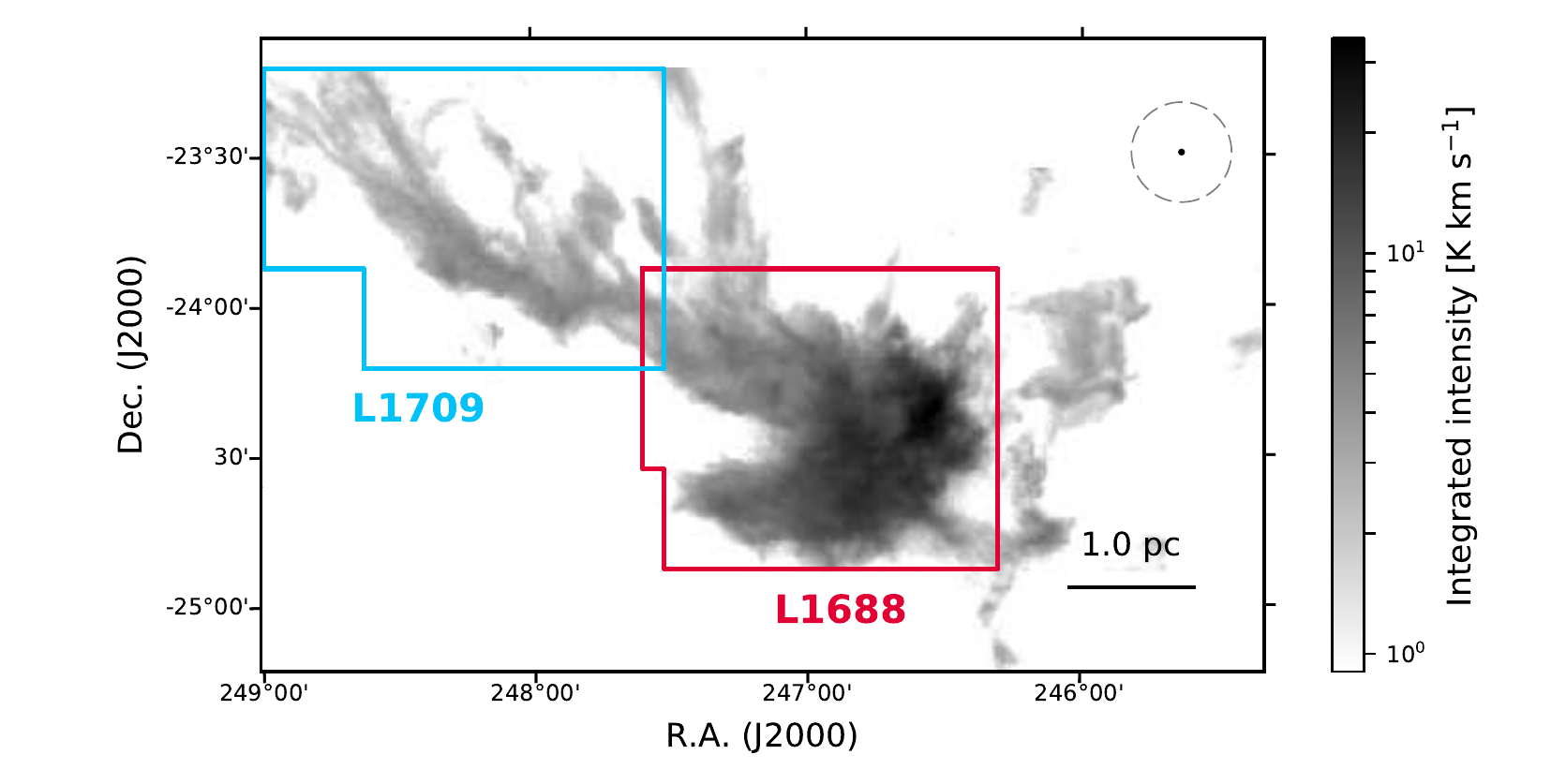}
\caption{Two selected sub-regions in the Ophiuchus cloud. The background image is the integrated intensity map of the $^{13}$CO line. The black dot in the gray dashed circle on the right-upper corner is the beam size. \label{fig_Oph13CO}}
\end{figure}

The Orion A cloud map is divided by galactic longitude, $l$, of 208$\degr$.25, 209$\degr$.75, 211$\degr$.75, and 213$\degr$.75. From the north to the south, the sub-regions are referred to as the ISF, Tail-N, Tail-S, and L1647-S regions, respectively (see Figure \ref{fig_Ori13CO}). Table \ref{tbl_D_Ori} shows an average distance to each sub-region. We also divide the Ophiuchus cloud map into two representative sub-regions, the L1688 and L1709 regions, which are two distinct parts with different system velocities \citep{Lor89b}. Figure \ref{fig_Oph13CO} exhibits the sub-regions in the Ophiuchus cloud. We adopt the same distance to these sub-regions as we adopt for the full cloud \citep[137~pc; ][]{Ort17}.

The scaling relations for the sub-regions within the Orion A and Ophiuchus clouds are presented in the left and right panels of Figure \ref{fig_LSdiag_subregion}, respectively. The $N_\mathrm{sig}$, $p_\mathrm{var}$, best-fit $\delta v_0$ and $\alpha$ values are summarized in Table \ref{tbl_PCA_Ori}. The scaling relations for the entire Orion A and Ophiuchus clouds are also overlaid in each panel. The scaling relations for the sub-regions have different $\delta v$ values for a given $L$, where the relation for the entire cloud seems to be an average of those for the sub-regions \citep{Bru99}. PCA results for the Orion A cloud show that the $\delta v$ for a given $L$ from the northern sub-region tends to be greater than that for the southern sub-region. The best-fit $\alpha$ values for the sub-regions are similar to each other ($\sim$ 0.7) except that for the ISF region. The ISF region has the largest $\delta v$ for a given $L$ and the highest $\alpha$ value of 0.93. In the Ophiuchus cloud, the $\delta v$ for a given $L$ for the L1688 region is higher than that for the L1709 region. The difference of $\delta v$ in the Ophiuchus cloud is relatively smaller than that in the Orion A cloud.
 
\subsection{PCA Results for Different Density Tracers} \label{PCA_lines}
The ISF and L1688 regions are the most active star-forming regions in the Orion A and Ophiuchus clouds, respectively. Also, the observed molecular lines in these regions are strong enough to provide good signal-to-noise ratios (Paper 1). The active star formation and high signal-to-noise ratios of the observed lines make the ISF and L1688 regions ideal to study the relationship between turbulence and star formation. We thus apply PCA to the observed line maps, which trace different density environments, toward the ISF and L1688 regions. Among the observed lines, the HCN and N$_2$H$^+$ lines have hyperfine components. PCA can consider these hyperfine components as different velocity components of a cloud. Therefore, only the $^{13}$CO, C$^{18}$O, HCO$^+$, and CS lines are used for the PCA analyses. 

\begin{figure}
\epsscale{1.0}
\plottwo{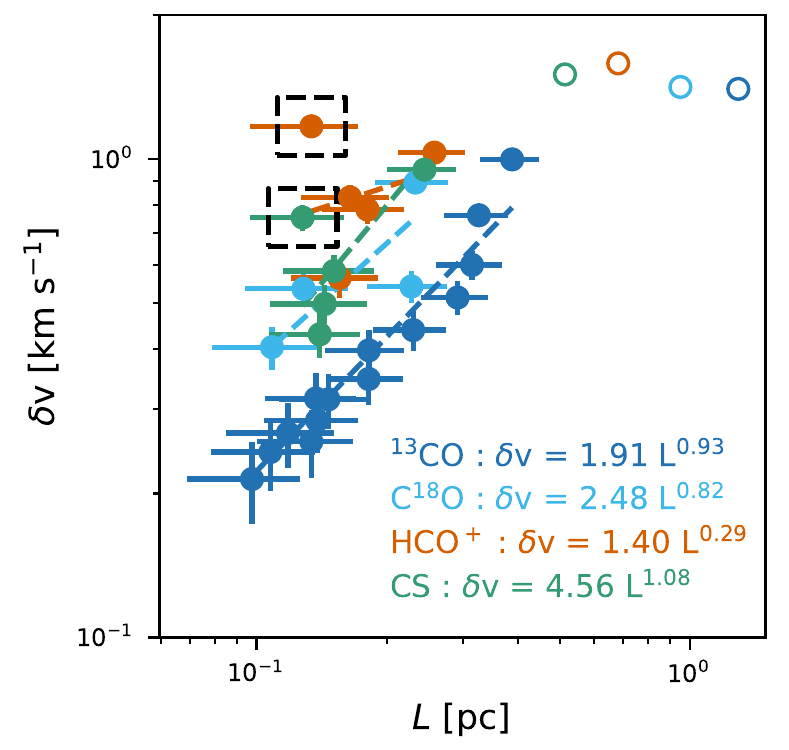}{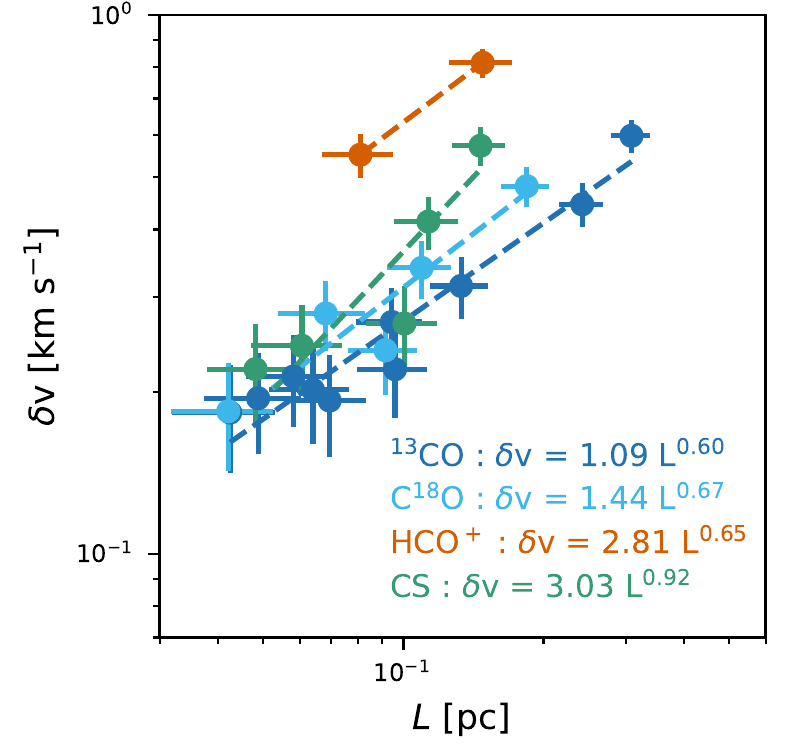}
\caption{The scaling relations for the $^{13}$CO, C$^{18}$O, HCO$^+$, and CS lines in the ISF (left) and L1688 (right) regions. The black dashed boxes in the left panel indicate the 6th PC of the HCO$^+$ line and the 4th PC of the CS line. \label{fig_LSdiag_lines}}
\end{figure}

Figure \ref{fig_LSdiag_lines} shows the scaling relations for the observed lines in the ISF and L1688 regions. The $N_\mathrm{sig}$, $p_\mathrm{var}$, and power-law fits are summarized in Table \ref{tbl_PCA_ISF}. The scaling relations for $^{13}$CO are identical to those presented in Figure \ref{fig_LSdiag_subregion}. Except for the $^{13}$CO line, the observed lines show only a few PCs, which are highly scattered within a narrow range of $L$ in Figure \ref{fig_LSdiag_lines}, resulting in large uncertainties in the $\alpha$ and $\delta v_0$ values. 

The scaling relations for C$^{18}$O, HCO$^+$, and CS have higher $\delta v$ values for a given $L$ compared to that of $^{13}$CO. In the Orion A cloud, the scaling relations for C$^{18}$O, HCO$^+$ and CS are clearly separate from that of $^{13}$CO. For the HCO$^+$ and CS lines, there is one PC that has an even higher $\delta v$ and departs from the power-law trends expected from the other components: one is the 6th PC of HCO$^+$ and the other is the 4th PC of CS. These PCs are marked with the black dashed boxes in the left panel of Figure \ref{fig_LSdiag_lines}. We discuss these components in more detail in Section \ref{sec_KL}. In the Ophiuchus cloud, the $\delta v$ values at a given $L$ for C$^{18}$O and CS are slightly higher than that of $^{13}$CO. The PCA result from the HCO$^+$ map clearly shows a higher $\delta v$ for a given $L$, compared to the result from the $^{13}$CO map.

\section{Discussion} \label{Sec_discs}
In this section, we investigate how scaling relations can be affected by star formation activity. We focus on the results for the Orion A cloud since it has both massive \citep{Ike07,Nak12} and low-mass \citep{All08} star-forming environments, and the scaling relations show the distinctively large $\delta v$ differences (the left panels of Figures \ref{fig_LSdiag_subregion} and \ref{fig_LSdiag_lines}).

\subsection{Difference in $\delta v$ at a Given $L$} \label{Sec_feedback}
\citet{Hey06} found a difference in $\delta v_0$ between the sub-regions within the Rosette molecular cloud. They suggested that the interaction between the H II region and surrounding clouds may inject energy and increase $\delta v_0$. This result implies that $\delta v$ can be affected by star formation activities within the sub-regions.

The left panel of Figure \ref{fig_LSdiag_subregion} shows that the $\delta v$ at a given $L$ decreases toward the southern part of the Orion A cloud, which is suggestive of effects of dynamical conditions. In Orion A, massive star formation occurs in the northern part \citep{Gre13}, while only low-mass star formation occurs in the southern part \citep{All08}. We explore the relation between star formation activity and $\delta v$ difference in more detail using the number density of Class 0/I young stellar objects (YSOs) and flat-spectrum sources (embedded protostars).

\begin{figure}
\epsscale{0.5}
\plotone{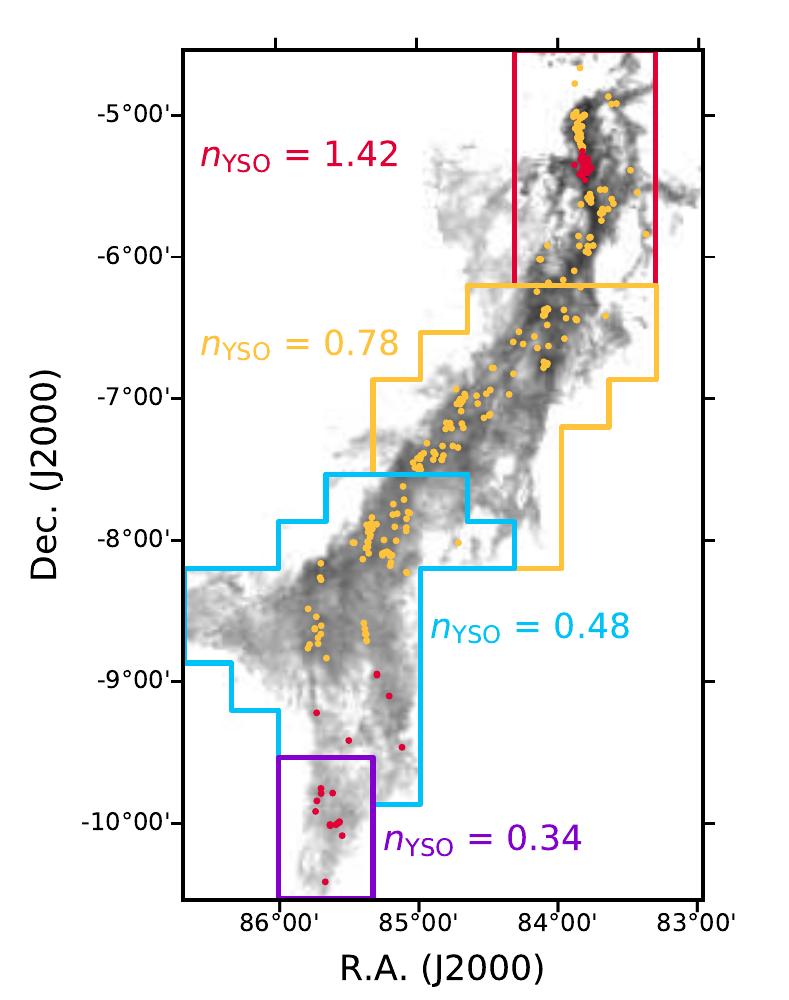}
\caption{Positions of the Class 0/I young stellar objects (YSOs) and flat-spectrum sources (the embedded protostars) in the Orion A cloud. The background image presents the integrated intensity map of the $^{13}$CO line. The embedded protostars identified with \textit{Herschel} observations \citep{Fur16} are represented by the yellow dots. Because \citet{Fur16} missed protostars near the Orion nebular and L1647, The YSO catalogue identified using \textit{Spitzer} space telescope \citep{Meg12} is adopted for these regions (red dots). The number density of the embedded protostars ($n_\mathrm{YSO}$) for each sub-region is summarized on the map. The unit of $n_\mathrm{YSO}$ is pc$^{-2}$. \label{fig_Ori_YSO}}
\end{figure}

For this analysis, we adopted the catalog of embedded protostars identified with \textit{Herschel} \citep{Fur16} and \textit{Spitzer} space telescope \citep{Meg12} observations. Figure \ref{fig_Ori_YSO} presents the distribution of protostars in the Orion A cloud. Table \ref{tbl_YSO_Ori} shows the number ($N_\mathrm{YSO}$), number densities ($n_\mathrm{YSO}$), and total bolometric luminosity ($L_\mathrm{bol}$) of the embedded protostars in each sub-region. The protostellar number density decreases from the ISF to the Tail-S region, and the values in the L1647-S and Tail-S regions are similar. This trend resembles the $\delta v$ difference between the sub-regions. The total $L_\mathrm{bol}$ also decreases in the sub-regions from the north to the south. 

\begin{figure}
\epsscale{0.5}
\plotone{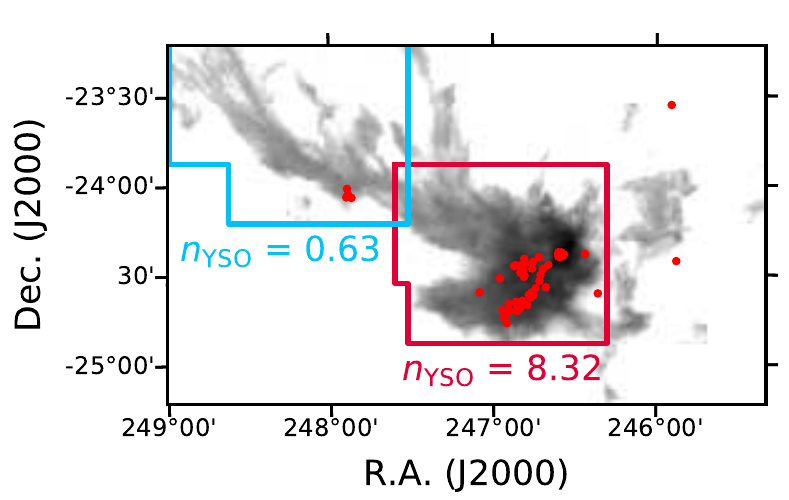}
\caption{The same as Figure \ref{fig_Ori_YSO} but for the Ophiuchus cloud. The red dots represent the embedded protostars identified with \textit{Spitzer} observations \citep{Dun15}. \label{fig_Oph_YSO}}
\end{figure}

We also investigated the $n_\mathrm{YSO}$ and total $L_\mathrm{bol}$ for the L1688 and L1709 regions. The \textit{Spitzer} YSO catalog produced by \citet{Dun15} was adopted to investigate the distribution of YSOs in the Ophiuchus cloud (see Figure \ref{fig_Oph_YSO} and Table \ref{tbl_YSO_Ori}). Similar to those for the Orion A cloud, both $n_\mathrm{YSO}$ and the total $L_\mathrm{bol}$ in the L1688 region are also higher than those for the L1709 region. Similar variation trends between the $\delta v$ of the PCA scaling relations and the number density and total luminosity of the embedded protostars in the sub-regions imply that the turbulent velocity difference is affected by the local star formation activities.

Figure \ref{fig_LSdiag_lines} shows that the $\delta v$ for a given $L$ also varies depending on the observed lines. At a certain $L$, the $\delta v$ for C$^{18}$O, HCO$^+$, and CS are higher than that for $^{13}$CO. The C$^{18}$O, HCO$^+$, and CS lines are optically thinner than the $^{13}$CO line because of their lower abundances \citep{Wil94,Lee98} or higher critical densities \citep{Ung97}. Consequently, these lines preferentially trace the dense environment compared to the $^{13}$CO line. Therefore, the high-$\delta v$ values for C$^{18}$O, HCO$^+$, and CS implies that the dense gas may be more turbulent than the diffuse gas in the ISF region. These results are remarkable because the high column density regions in both clouds have broader $^{13}$CO linewidth than the envelope regions. Also, the linewidth of $^{13}$CO is generally broader than that of C$^{18}$O, HCO$^+$, and CS (Paper 1).

Another important point is that the scaling relation for HCO$^+$ tends to have the largest $\delta v$ in both clouds. This is likely because the HCO$^+$ line is mainly detected in the regions where stars actively form (Paper 1). \citet{Shi03} derived the Type-2 ({\it single-tracer, multicloud}) linewidth-size relation using C$^{34}$S $J=$5$-$4 for 51 high-mass star-forming cores. They found much broader line widths at a given size than those for the Type-1 ({\it multitracer, multicloud}) linewidth-size relations of ``high-mass" and ``low-mass" cores derived by \citet{Cas95}. Certain molecular transitions preferentially trace gas that is much more turbulent than gas in other parts of the cloud. However, our result that higher $\delta v$s are derived from higher density tracers requires some discussion because it is conventionally accepted that the turbulence dissipates in dense environments, thereby resulting in narrower linewidths \citep{Mye83,Nak98,Goo98}. 

\subsection{Effect of the Orion KL outflows on the PCA} \label{sec_KL}
In the ISF, there are energetic features that originate from active star formation such as high-velocity outflows from Orion KL in OMC-1. Broad wing structures, produced by the Orion KL outflows, clearly appear in the $^{13}$CO, HCO$^+$ and CS lines and marginally in the C$^{18}$O line (Paper 1). High-velocity outflows may provide the key to understanding the PCA results derived from the $^{13}$CO maps, where we find $\delta v$ varies with the $n_\mathrm{YSO}$ and total $L_\mathrm{bol}$ of the sub-region (see Section \ref{Sec_feedback}). 

\begin{figure}
\epsscale{1.0}
\plotone{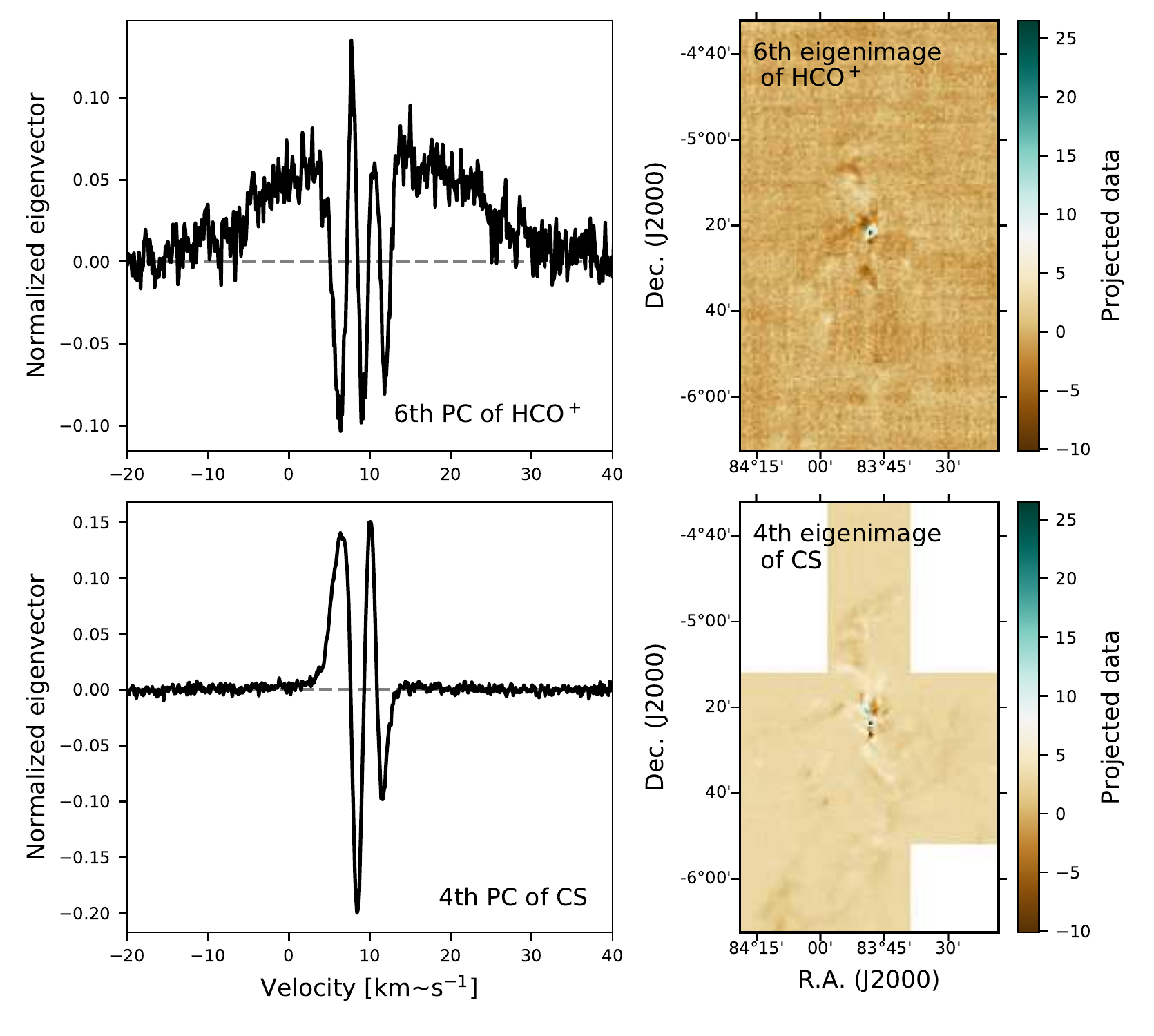}
\caption{Eigenvectors (left) and eigenimages (right) for the PCs that are affected by the Orion KL outflows. The eigenvector and eigenimage for the 6th PC of HCO$^+$ are presented in the top panels, and those for the 4th PC of CS are presented in the bottom panels. \label{fig_comp_profile}}
\end{figure}

The scaling relations for the observed lines in the ISF region show that there are two PCs that have high-$\delta v$ values at a given $L$ compared to the other PCs (see the black dashed boxes in Figure \ref{fig_LSdiag_lines}): one is the 6th PC of HCO$^+$ and the other is the 4th PC of CS. Figure \ref{fig_comp_profile} shows the eigenvectors and eigenimages of the 6th PC of HCO$^+$ and 4th PC of CS. The HCO$^+$ eigenvector exhibits high-velocity wing structures that extend from $-$20 to 35~km~s$^{-1}$. Its eigenimage also shows a large positive projection near Orion KL, which indicates that this component mainly describes a velocity difference near Orion KL. These features imply that the high-$\delta v$ of the 6th PC of HCO$^+$ probably originates from the Orion KL outflows, and the localized star formation activities increase the $\delta v$ value of a specific PC. For the CS line, the eigenvector does not clearly show the broad wing structures although its eigenimage presents large positive and negative projection values near Orion KL. These features may be due to the weak intensities of the broad wing structures compared to those for the HCO$^+$ line (Paper 1). 

We derived the scaling relations for the observed lines in the ISF excluding the spectra affected by the Orion KL outflows from the PCA. Since the outflow features are only detected near Orion KL, we can exclude the outflow contamination by removing the line spectra within a circular area that is centered at Orion KL ($\alpha_\mathrm{J2000}$=5$^\mathrm{h}$35$^\mathrm{m}$14$^\mathrm{s}$.16, $\delta_\mathrm{J2000}$=$-$5$\degr$22$^\mathrm{m}$21$^\mathrm{s}$.5) with a diameter of 5$\arcmin$. Figure \ref{fig_LS_ISF_mask_KL} shows the scaling relations without the Orion KL outflows. In these scaling relations, the PCs with the high-$\delta v$ values (the 5th PC of HCO$^+$ and 3rd PC of CS in the original results; see Figure \ref{fig_LSdiag_lines}) disappear. This result confirms that the Orion KL outflows indeed enhance the $\delta v$ values of the 6th PC of HCO$^+$ and 4th PC of CS.

\subsection{Steep Slope and Low-$\delta v$ of $^{13}$CO in the ISF} \label{Sec_large_scale}
\begin{figure}
\epsscale{1.0}
\plottwo{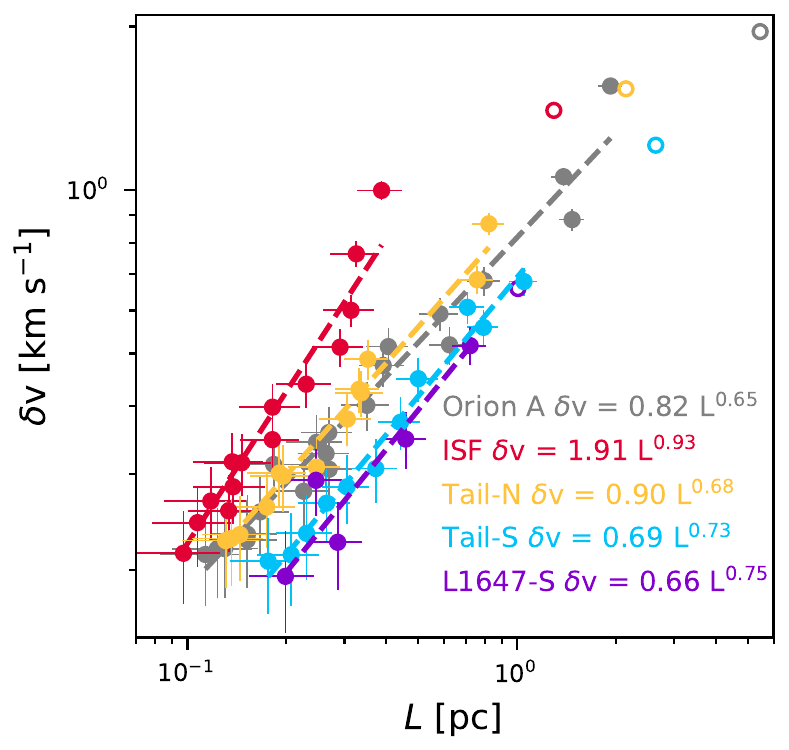}{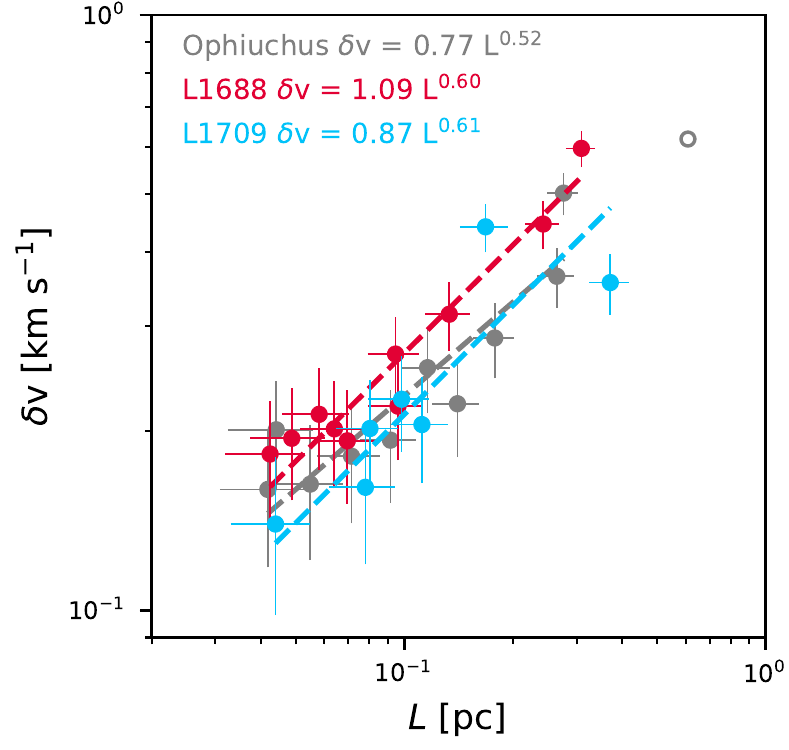}
\caption{The scaling relations for the $^{13}$CO line in the sub-regions of the Orion A (left) and Ophiuchus (right) clouds. The best-fit power-law relations are exhibited by the dashed lines. In each panel, the scaling relation for the entire cloud is presented in gray. \label{fig_LSdiag_subregion}}
\end{figure}

\begin{figure}
\epsscale{0.5}
\plotone{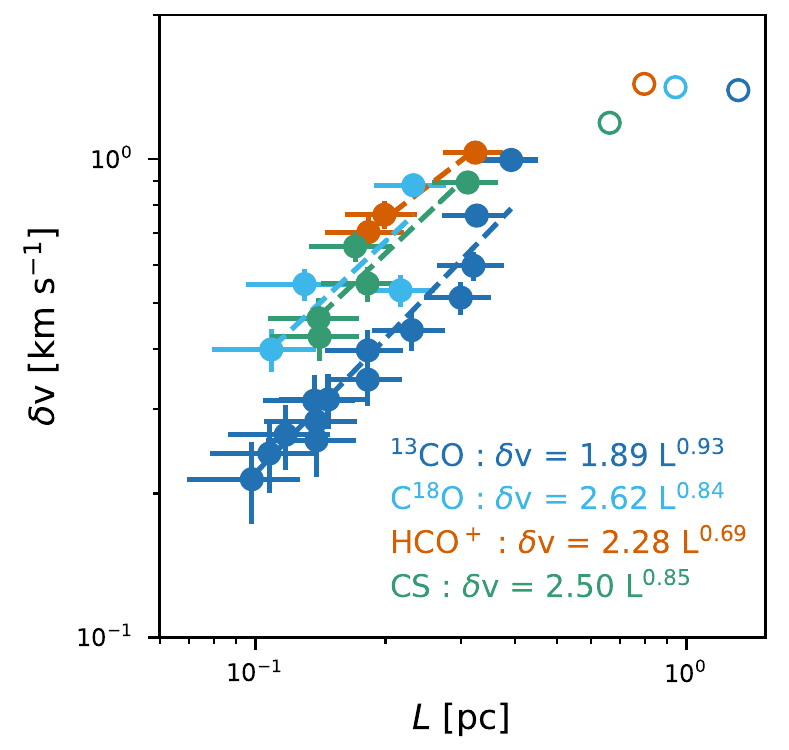}
\caption{The scaling relations for the $^{13}$CO, C$^{18}$O, HCO$^+$, and CS lines in the ISF region after excluding the Orion KL outflows. The PCA results are derived from the ISF region except near Orion KL (a circular area centered on Orion KL with a diameter of 5$\arcmin$). \label{fig_LS_ISF_mask_KL}}
\end{figure}

The scaling relation for $^{13}$CO in the ISF region has a higher $\alpha$ than those for the other sub-regions (see Figure \ref{fig_LSdiag_subregion}). This steep slope does not appear to be caused by the Orion KL outflows because it still appears in the scaling relation after the Orion KL region is removed (see Figure \ref{fig_LS_ISF_mask_KL}). The large-$\alpha$ could be attributed to a steep increase of $\delta v$ at $L$ of about 0.3~pc. 

\begin{figure}
\epsscale{0.5}
\plotone{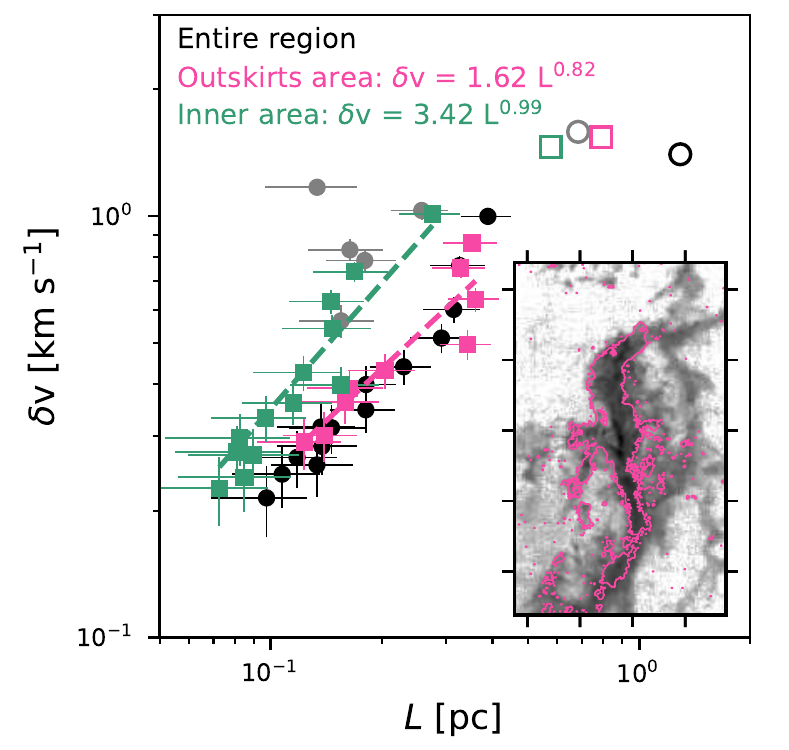}
\caption{The scaling relation for the $^{13}$CO line in the outskirts (the pink squares) and inner area (the green squares) of the ISF region. The boundary of the outskirts and inner areas (the pink contour) is defined where the HCO$^+$ line is detected over the 5-$\sigma$ level. The original scaling relations for $^{13}$CO (the black circles) and HCO$^+$ (the gray circles) are also overlaid for a comparison. \label{fig_LS_ISF_mask_out}}
\end{figure}

We mask the inner area of the ISF region, where the HCO$^+$ line is detected over 5-$\sigma$, from the $^{13}$CO map (see the pink contour in Figure \ref{fig_LS_ISF_mask_out}) to assess the origin of the high $\delta v$ values at $L$ $\geqq$ 0.3~pc. The PCA result for the masked $^{13}$CO data (the pink squares) represents the gas motion in the remaining outskirts of the ISF. Figure \ref{fig_LS_ISF_mask_out} shows that the drastic change of $\delta v$ at $L$ $\sim$ 0.3~pc still exists. This velocity excess could still be related to the Orion KL outflow if it is produced by turbulence excited by the outflow expansion. \citet{Off18} and \citet{Off17} showed that stellar winds and outflows can excite magnetosonic waves, which propagate away from where the feedback is launched, thereby enhancing turbulence in regions removed from where the feedback is produced. Alternatively, the steep increase in $\delta v$ at $L$ $\geqq$ 0.3~pc could be caused by large-scale motion of the filamentary structure rather than by local star formation activity, such as outflows and jets.

We also apply the PCA to the inner area of the $^{13}$CO map (the green squares), where the HCO$^+$ line is detected. The derived scaling relation shows velocity differences greater than those derived from only the outskirt area. However, they are as large as those derived from the HCO$^+$ map (the gray circles). This result implies that the $^{13}$CO line also traces the gas affected by the active star formation in ISF as the HCO$^+$ line does. However, in the PCA for the entire ISF region, these high $\delta v$ values of the inner area would be diluted because of the contribution of the outskirts. 

\subsection{Large-scale motion of the ISF}
Large-scale systematic motion, such as rotation and shear, affects the PCA result, and thus, we should consider its effect to determine the properties of turbulence \citep{Bur00,Fed16}. The steep increase of $\delta v$ at 0.3~pc, which was discussed in Section \ref{Sec_large_scale}, probably results from a large-scale velocity gradient across the ISF. Therefore, we subtract the large-scale motion from the original velocity field to check the effect of the outskirts of the ISF on the PCA (see magenta symbols in Figure \ref{fig_LS_ISF_mask_out}).

\begin{figure}
\epsscale{1.0}
\plotone{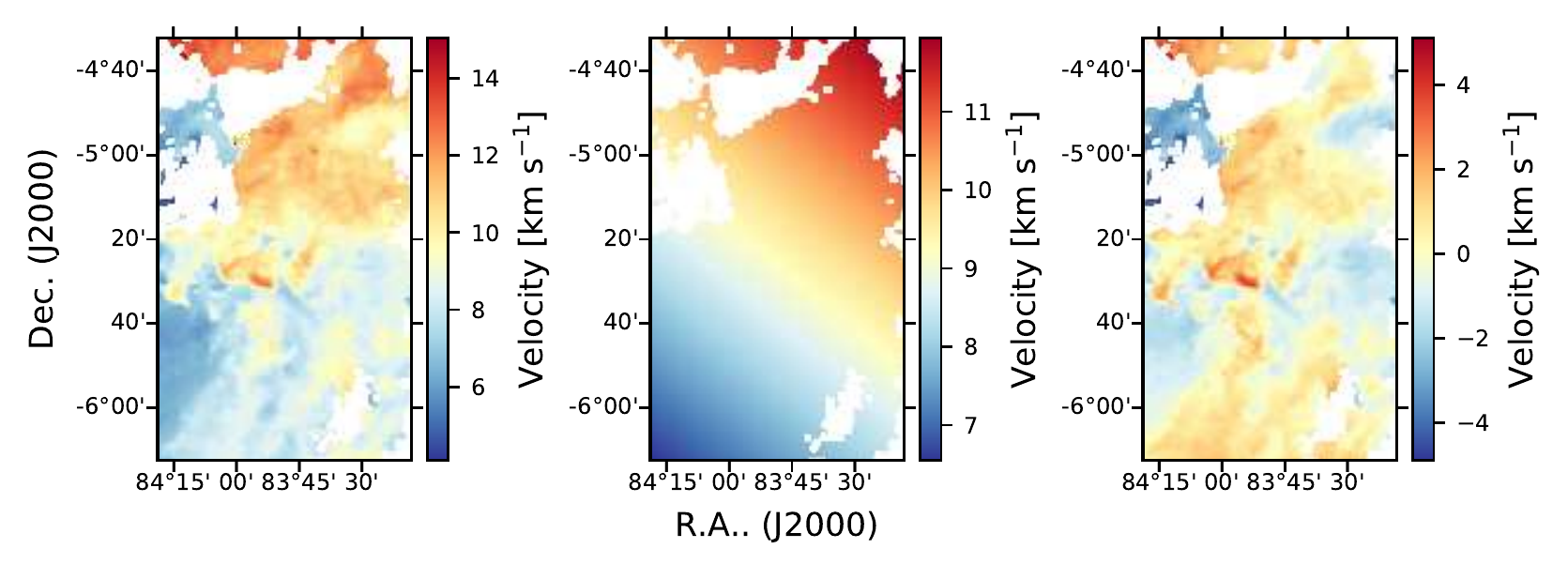}
\caption{Left: Intensity weighted velocity (moment 1) map of the $^{13}$CO line in the ISF region (Paper 1). Middle: large-scale velocity gradient in the ISF region. The large-scale gradient is isolated from the turbulent velocity field via fitting the moment 1 map with a plane. Right: Moment 1 map for the remaining turbulent velocity field. \label{fig_ISF_13CO_vsub}}
\end{figure}

\citet{Fed16} subtracted the large-scale systematic motion of the central molecular zone cloud to obtain the velocity distribution function of turbulence. They fit the intensity weighted velocity (moment 1) map with a plane to isolate the systematic velocity gradient of the cloud from the turbulent velocity field. We apply the same method to the $^{13}$CO data of the ISF. We fit the moment 1 map of the ISF, which is adopted from Paper 1, with a plane. Subsequently, the fitted velocity gradient across the ISF is subtracted from the original $^{13}$CO cube data (see Figure \ref{fig_ISF_13CO_vsub}).

\begin{figure}
\epsscale{1.0}
\plottwo{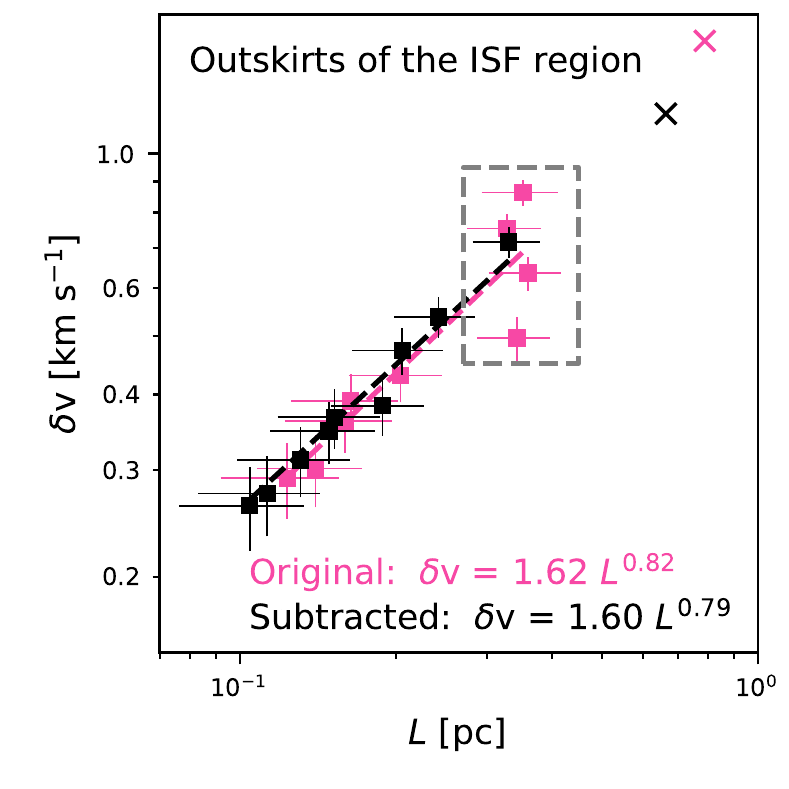}{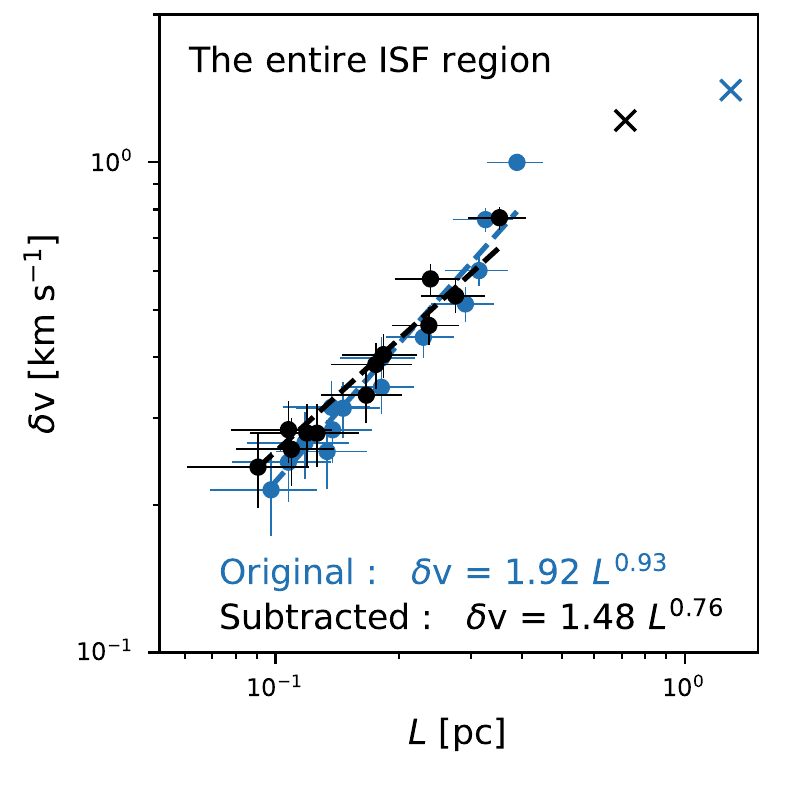}
\caption{The scaling relations for the $^{13}$CO line in the outskirts (left) and entire ISF region (right) after the subtraction of the large-scale velocity gradient. In each panel, the scaling relation for the original data is overlaid for comparison. The gray dashed box in the left panel indicates the steep increase of $\delta v$ at 0.3~pc. \label{fig_LS_ISF_vsub}}
\end{figure}

Figure \ref{fig_LS_ISF_vsub} presents the PCA results after the subtraction of the systematic velocity gradient. The steep increase of $\delta v$ disappears after the velocity gradient was subtracted (the gray dashed box). Also, all ($\delta v$, $L$) points are well aligned and can be well fit with a single power-law. When we apply the PCA to the entire ISF region, the $\alpha$ value decreases from 0.93 to 0.76 after the subtraction. These results strongly suggest that the steep increase in $\delta v$ originates from the large-scale motion of the ISF.

\section{Summary} \label{Sec_sum}
To study the relation between turbulence and star formation in molecular clouds, we observed the Orion A and Ophiuchus clouds in six different molecular lines as one of TRAO-KSPs, ``mapping Turbulent properties In star-forming MolEcular clouds down to the Sonic scale (TIMES)" (Paper 1). We investigated the properties of turbulence traced by $^{13}$CO $J$=1$-$0, C$^{18}$O $J$=1$-$0, HCO$^+$ $J$=1$-$0, and CS $J$=2$-$1 by applying a statistical method, PCA. The main results of the analyses are summarized as follows. 

\begin{enumerate}
\item The uniform distribution of $T_\mathrm{rms}$ allows us to access the gas motion on small scales using PCA.
\item For the entire Orion A and Ophiuchus clouds, PCA scaling relations for the $^{13}$CO line have $\alpha$ and $\delta v_0$ values that are consistent with those of the universality of turbulence proposed by \citet{Hey04}.
\item When we apply PCA to the $^{13}$CO data in the sub-regions of the Orion A and Ophiuchus clouds, the $\delta v$ for a given $L$ varies depending on the sub-regions. In each cloud, the $\delta v$ for a given $L$ is generally higher in sub-regions that are more actively star-forming. 
\item The variation in $\delta v$ is related to the $n_\mathrm{YSO}$ and total $L_\mathrm{bol}$ of the sub-regions. Meanwhile, localized phenomena, such as the Orion KL outflows, can change the $\delta v$ of a specific PC.
\item The scaling relations of the observed lines in the ISF and L1688 regions show that the $\delta v$ of C$^{18}$O, HCO$^+$, and CS are generally higher than that of $^{13}$CO. This implies that the dense gas is more turbulent than the diffuse gas in these regions, probably due to energy input from active star formation in the dense regions. 
\item The scaling relation of the $^{13}$CO line only in the dense inner part of the ISF shows $\delta v$ similar to that derived from the HCO$^+$ line. However, this high $\delta v$ would be diluted in the PCA of the entire ISF region because the scaling relation is more affected by the larger outskirt area.
\item The scaling relation for the $^{13}$CO line in the ISF region has a steep increase of $\delta v$ at $L$ of about 0.3~pc resulting in a large $\alpha$. This increase in $\delta v$ may be due to magnetosonic waves excited by Orion KL or due to large-scale motion of the filamentary cloud. 
\end{enumerate}

\section*{Acknowledgments}
This work was supported by the National Research Foundation of Korea (NRF) grant funded by the Korea government (MSIT) (grant number 2021R1A2C1011718).

\clearpage

\begin{figure}
\epsscale{0.5}
\plotone{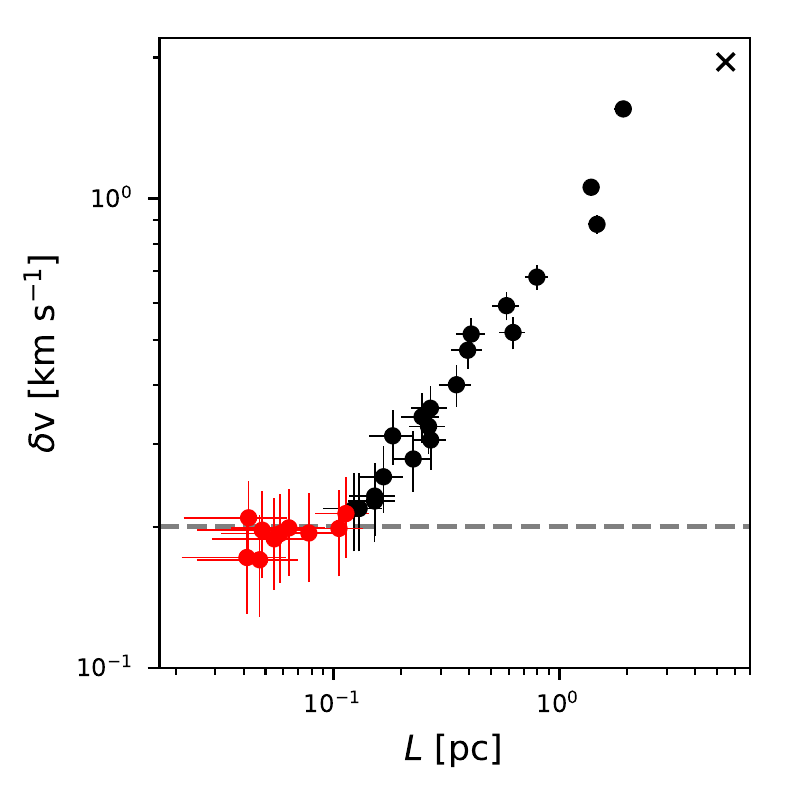}
\caption{An example of the PCA result for the $^{13}$CO line in the Orion A cloud. The black cross represents the ($\delta v$, $L$) point for the 2nd PC, and the black circles represent those from the 3rd to 22nd PCs. The ($\delta v$, $L$) points from the 23rd to the 32nd PCs are marked with the red circles. The gray dashed line indicates $\delta v$ of 0.2~km~s$^{-1}$, and the error bars show the 1-$\sigma$ error ranges. Note that the ($\delta v$, $L$) points of the 1st PC is omitted. \label{fig_LSD_example}}
\end{figure}

\begin{deluxetable}{lcc}
\tablecolumns{3}
\tabletypesize{\scriptsize}
\tablecaption{Distance to the Orion A cloud \label{tbl_D_Ori}}
\tablewidth{0pt}
\tablehead{
  \colhead{$l$ bin\tablenotemark{a}}&\colhead{$\bar{d}_\mathrm{YSOs}$\tablenotemark{a}}&\colhead{$\bar{d}_\mathrm{Reg}$\tablenotemark{b}}\\
  \colhead{[degree]}&\colhead{[pc]}&\colhead{[pc]}
}
\startdata
\sidehead{ISF}
208.0 - 209.0& 391$\pm$24& \\
208.5 - 209.5& 393$\pm$25& 392\\
209.0 - 210.0& 393$\pm$22& \\
\sidehead{Tail-N}
209.5 - 210.5& 390$\pm$26& \\
210.0 - 211.0\vspace{-0.2cm}& 395$\pm$30& \\
\vspace{-0.2cm}& & 399\\
210.5 - 211.5& 401$\pm$30& \\
211.0 - 212.0& 409$\pm$32& \\
\sidehead{Tail-S}
211.5 - 212.5& 417$\pm$44& \\
212.0 - 213.0\vspace{-0.2cm}& 423$\pm$46& \\
\vspace{-0.2cm}& & 431\\
212.5 - 213.5& 435$\pm$36& \\
213.0 - 214.0& 448$\pm$32& \\
\sidehead{L1647-S}
213.5 - 214.5\vspace{-0.2cm}& 461$\pm$40& \\
\vspace{-0.2cm}& & 464\\
214.0 - 215.0& 467$\pm$38& \\
\enddata
\tablenotetext{a}{From \citet{Gro18}}
\tablenotetext{b}{The arithmetic mean of the $\bar{d}_\mathrm{YSOs}$ values within each sub-region.}
\end{deluxetable}

\clearpage
\begin{deluxetable}{lccccc}
\tablecolumns{6}
\tabletypesize{\scriptsize}
\tablecaption{Parameters for the TRAO and FCRAO data \label{tbl_data_params}}
\tablewidth{0pt}
\tablehead{
  \colhead{Data}&	\colhead{Line}&	\colhead{Beam size}&	\colhead{Pixel size}&	\colhead{Velocity resolution}&	\colhead{$T_\mathrm{rms}$}\\
  \colhead{}&	\colhead{}&	\colhead{[$\arcsec$]}&	\colhead{[$\arcsec$]}&	\colhead{[km~s$^{-1}$]}&	\colhead{[K]}
}
\startdata
TRAO data&	$^{13}$CO $J=$1$-$0&	46&	20&	0.084&	0.228$\pm$0.021\\
FCRAO data&	$^{13}$CO $J=$1$-$0&	46&	23&	0.066&	0.244$\pm$0.058\\
\enddata
\end{deluxetable}

\begin{deluxetable}{lcccc}
\tablecolumns{5}
\tabletypesize{\scriptsize}
\tablecaption{The PCA results for the $^{13}$CO line for each cloud \label{tbl_PCA_clouds}}
\tablewidth{0pt}
\tablehead{
  \colhead{Cloud}&\colhead{$N_\mathrm{sig}$}&\colhead{$p_\mathrm{var}$}&\colhead{$\log_{10}$($\delta v_0$)}&\colhead{$\alpha$}
}
\startdata
Orion A&	22&	72.1&	-0.09$\pm$0.02&	0.65$\pm$0.05\\
Ophiuchus&	12&	90.3&	-0.12$\pm$0.13&	0.52$\pm$0.14\\
\enddata
\end{deluxetable}

\begin{deluxetable}{lcccc}
\tablecolumns{5}
\tabletypesize{\scriptsize}
\tablecaption{The PCA results for the sub-regions \label{tbl_PCA_Ori}}
\tablewidth{0pt}
\tablehead{
  \colhead{Region}&\colhead{$N_\mathrm{sig}$}&\colhead{$p_\mathrm{var}$}&\colhead{$\log_{10}$($\delta v_0$)}&\colhead{$\alpha$}
}
\startdata
\sidehead{In the Orion A cloud}
ISF		&   16& 84.3&  0.282$\pm$0.124& 0.933$\pm$0.182\\
Tail-N	&   15& 75.5& -0.047$\pm$0.046& 0.684$\pm$0.096\\
Tail-S	&   13& 63.8& -0.161$\pm$0.037& 0.731$\pm$0.112\\
L1647-S	&   7&  35.4& -0.181$\pm$0.103& 0.747$\pm$0.249\\
\sidehead{In the Ophiuchus cloud}
L1688   &   11& 96.5&  0.036$\pm$0.097& 0.604$\pm$0.109\\
L1709   &   8&  72.6& -0.062$\pm$0.191& 0.609$\pm$0.219\\
\enddata
\end{deluxetable}

\begin{deluxetable}{lcccccccc}
\tablecolumns{9}
\tabletypesize{\scriptsize}
\tablecaption{The PCA results for the observed lines \label{tbl_PCA_ISF}}
\tablewidth{0pt}
\tablehead{
  \colhead{}&\multicolumn{4}{c}{The ISF}&\multicolumn{4}{c}{L1688}\\
  \colhead{line}&\colhead{$N_\mathrm{sig}$}&\colhead{$p_\mathrm{var}$}&\colhead{$\log_{10}$($\delta v_0$)}&\colhead{$\alpha$}&\colhead{$N_\mathrm{sig}$}&\colhead{$p_\mathrm{var}$}&\colhead{$\log_{10}$($\delta v_0$)}&\colhead{$\alpha$}
}
\startdata
$^{13}$CO& 16& 84.3& 0.281$\pm$0.122& 0.932$\pm$0.179& 11& 96.5& 0.036$\pm$0.099& 0.604$\pm$0.111\\
C$^{18}$O& 6&  6.9&  0.374$\pm$2.198& 0.788$\pm$2.755& 6&  43.0& 0.159$\pm$0.232& 0.665$\pm$0.238\\
HCO$^+$&   7&  21.7& 0.161$\pm$5.494& 0.311$\pm$7.348& 3&  8.8&  0.450\tablenotemark{a}& 0.650\tablenotemark{a}\\
CS&        7&  55.7& 0.685$\pm$4.957& 1.110$\pm$6.165& 6&  45.5& 0.481$\pm$0.593& 0.920$\pm$0.574\\
\enddata
\tablenotetext{a}{For the HCO$^+$ line in the L1688 region, only two principal components are used to fit a power-law relation so that there is no estimated error.}
\end{deluxetable}

\begin{deluxetable}{lcccc}
\tablecolumns{5}
\tabletypesize{\scriptsize}
\tablecaption{Distribution of the embedded protostars \label{tbl_YSO_Ori}}
\tablewidth{0pt}
\tablehead{
  \colhead{Region}&\colhead{$N_\mathrm{YSO}$}&\colhead{Area\tablenotemark{a}}&\colhead{$n_\mathrm{YSO}$}&\colhead{Total $L_\mathrm{bol}$\tablenotemark{b}}\\
  \colhead{}&\colhead{}&\colhead{[pc$^2$]}&\colhead{[pc$^{-2}$]}&\colhead{[$L_\sun$]}
}
\startdata
\sidehead{In the Orion A cloud}
ISF		&111&78.0	&1.42& 973.9\\
Tail-N	&88	&113.2	&0.78& 722.0\\
Tail-S	&78	&163.5	&0.48& 106.1\\
L1647-S	&15	&43.7	&0.34& -\\
\sidehead{In the Ophiuchus cloud}
L1688   &50 &6.0    &8.32& 59.2\\
L1709   &4  &6.3    &0.63& 2.6\\
\enddata
\tablenotetext{a}{For the Orion A cloud, we adopted the distances presented in Table \ref{tbl_D_Ori} to derive the area for the sub-regions. For the Ophiuchus cloud, the distance of 137~pc is adopted \citep{Ort17}.}
\tablenotetext{b}{The total bolometric luminosities for the sub-regions in the Orion A and Ophiuchus clouds are derived from the YSO catalogs provided by \citet{Fur16} and \citet{Dun15}, respectively. Note that the protostars near OMC-1 and L1647 are not included in the catalogs.}
\end{deluxetable}
\clearpage

\appendix
\renewcommand\thefigure{\thesection.\arabic{figure}}
\renewcommand\thetable{\thesection.\arabic{table}}
\section{The selection of significant principal components} \label{App_scree}
\setcounter{figure}{0}
\setcounter{table}{0}
One important factor in the PCA is the number of significant PCs ($N_\mathrm{sig}$) that covers most of the variation in the data \citep{PCABook}. The scree plot method \citep{Cat66} is one of the standard methods to evaluate $N_\mathrm{sig}$ in multivariate statistics \citep{Zos90,PCABook,Kan05}. In this study, the significant PCs were selected by checking the variation of $\delta v$ (see Section \ref{Sec_pca}). Here, we apply the scree plot method to our PCA result for the $^{13}$CO line in the Orion A cloud to check the reliability of our method.

The scree plot \citep{Cat66} is a plot of the eigenvalue ($\lambda_\mathrm{i}$) against the order number (i). The scree plot can have elbow points, which separate the scree plot into two parts: `steep' and `shallow', which are located in to the left and right, respectively, of each elbow point. If the scree plot has a single elbow point, the left and right of the elbow point are labeled as `Cliff' and `Scree', respectively. It is possible to have multiple elbow points in a scree plot \citep{Cat66,PCABook}. In this case, the left of the first elbow point is the `Cliff' regime \citep{Zos90,Kan05}. The PCs in the `Cliff' regime are considered significant in many studies \citep{Cat66,Zos90,PCABook,Kan05}.  

\begin{figure}
\epsscale{1.0}
\plotone{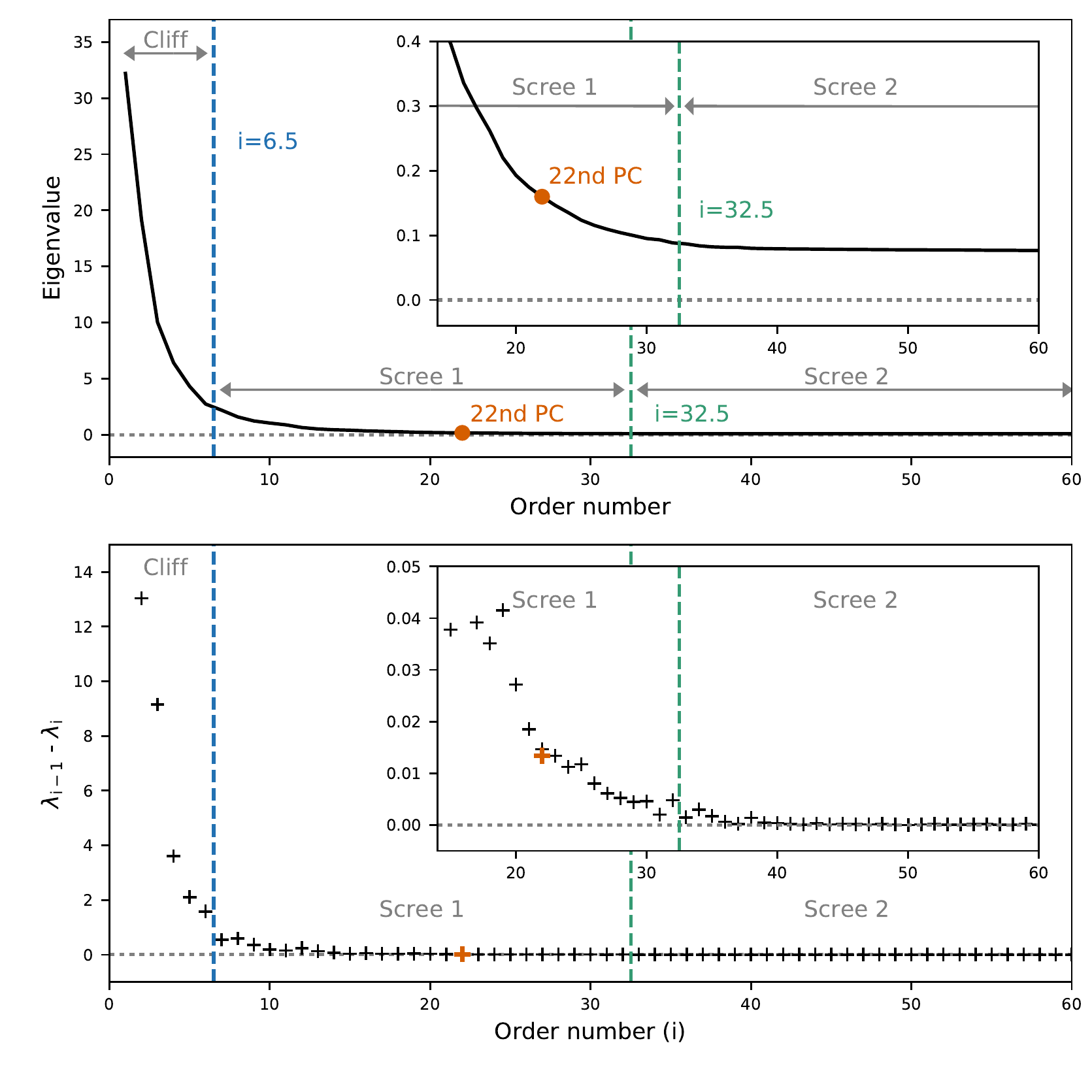}
\caption{Scree plot (top) and the variation of eigenvalue (i.e., $\lambda_\mathrm{i-1} - \lambda_\mathrm{i}$) as a function of order number (bottom) for the $^{13}$CO line in the Orion A cloud. The insets are the zoomed-in views for i from 14 to 60. The blue dashed lines indicate the boundary between the `Cliff' and `Scree 1' regimes (i=6.5), and the green dashed lines indicate the boundary between the `Scree 1' and `Scree 2' regimes (i=32.5). The 22nd PC, which is the largest order PC selected by our method, is marked with the orange symbols. \label{fig_scree}}
\end{figure}

Figure \ref{fig_scree} shows the scree plot for the $^{13}$CO line in the Orion A cloud. We also present the variation of $\lambda_\mathrm{i}$ (i.e., $\lambda_\mathrm{i-1} - \lambda_\mathrm{i}$) as a function of i, which is frequently used as a criterion to identify significant PCs \citep{PCABook}. The first elbow point is in between $\mathrm{i}=6$ and $7$ (the blue dashed line) since $\lambda_\mathrm{i-1} - \lambda_\mathrm{i}$ remains relatively constant beyond the 6th PC \citep{PCABook}. It is difficult to determine whether additional elbow points exist because the scree plot changes smoothly. We thus consider the left and right of the first elbow point as a `Cliff' and `Scree' regimes, respectively. The $\lambda_\mathrm{i}$ continuously decrease in the `Scree' regime up to the 32nd order. Beyond the 32nd PC, $\lambda_\mathrm{i}$ remains constant, and $\lambda_\mathrm{i-1} - \lambda_\mathrm{i}$ is close to zero. We thus divide the `Scree' regime by the 32nd PC into `Scree 1' and `Scree 2'; the `Scree 1' covers from i=7 to 32, and the `Scree 2' is beyond i=32.

Surprisingly, $N_\mathrm{sig}$ from the scree plot is six, which is much smaller than that from our method ($N_\mathrm{sig}$=22; the orange symbols in Figure \ref{fig_scree}). The percentage of total variation ($p_\mathrm{var}$; see Equation \ref{equ_pvar}) for the PCs that were selected with the scree plot and our methods are 63 and 72~\%, respectively. If we consider all components within the `Cliff' and `Scree 1' regimes (PCs from 1st to 32nd), $p_\mathrm{var}$ is 73~\%, which is not significantly different from the result of our selection method. These results imply that the scree plot method is a more conservative method for evaluating $N_\mathrm{sig}$, but our method selects all PCs that are significant to describe the most variation caused by turbulence, limiting the effect of noise.

\begin{figure}
\epsscale{1.0}
\plotone{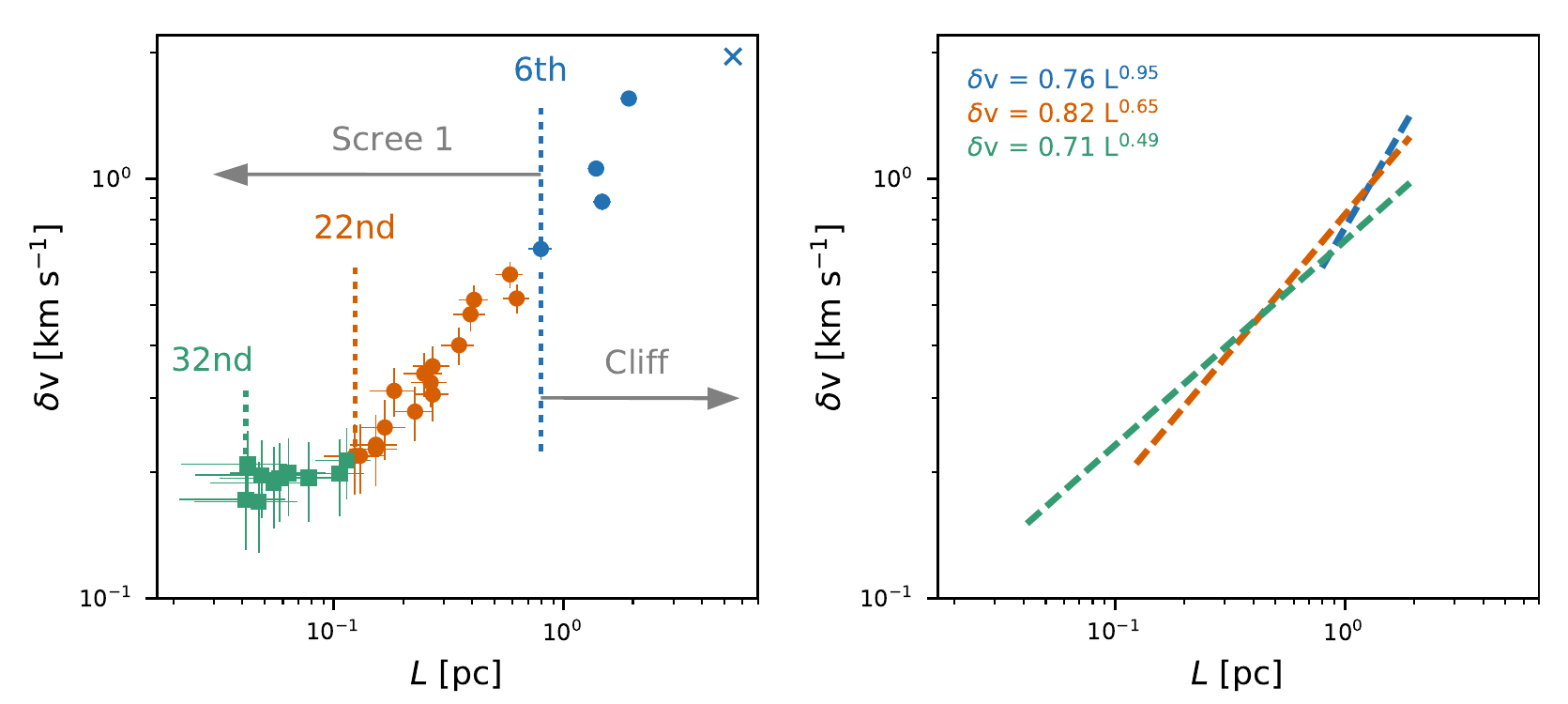}
\caption{Left: the ($\delta v$, $L$) points for the PCs in the `Cliff' and `Scree 1' regimes (up to the 32nd PC). The 1st PC is omitted, and the 2nd PC is presented by the blue cross symbol. The PCs within the three groups, which are divided by the 6th and the 22nd PCs (see the text), are exhibited in different colors. Right: The scaling relation for the PCs from the 3rd to the last component in each of the three groups. The best-fit relations are summarized in the upper-left corner. \label{fig_scree_LSD}}
\end{figure}

The left panel of Figure \ref{fig_scree_LSD} shows the ($\delta v$, $L$) points of the PCs within the `Cliff' and `Scree 1' regimes. Note that the 1st PC is omitted, and the 2nd PC is marked with the cross symbol as marked in Figure \ref{fig_LSD_example}. We divide the PCs into three groups: (1) from the 3rd to the 6th (in the `Cliff' regime; the blue symbols), (2) from the 7th to the 22nd (the PCs in the `Scree 1' regime and selected via our method; the orange symbols), and (3) from the 23rd to the 32nd (the rest PCs in the `Scree 1' regime; the green symbols). The best-fit scaling relation for the ($\delta v$, $L$) points from the 3rd to the last component in each group is presented in the right panel of Figure \ref{fig_scree_LSD}. The $N_\mathrm{sig}$, $p_\mathrm{var}$, best-fit $\delta v_0$ and $\alpha$ values are summarized in Table \ref{tbl_PCA_scree}.

\begin{figure}
\epsscale{1.0}
\plotone{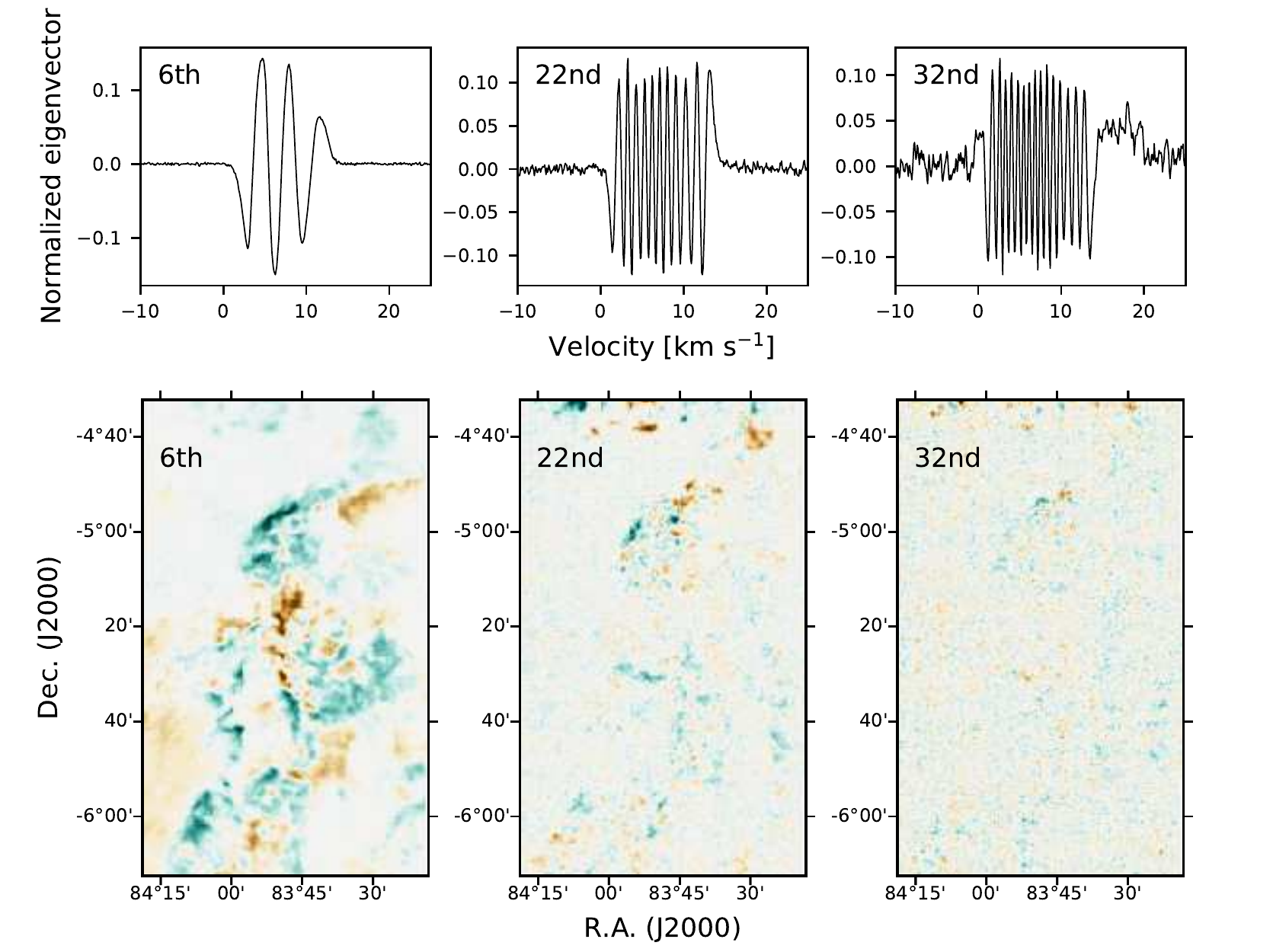}
\caption{Eigenvectors (top) and eigenimages (bottom) for the last component in each of the three groups defined in Appendix \ref{App_scree}. \label{fig_scree_img}}
\end{figure}

Beyond the 22nd PC, $\delta v$ converges into about 0.2~km~s$^{-1}$, and consequently $\alpha$ decreases as more PCs are included in the power-law fitting. The eigenvector and eigenimage of the last-order PC in each group (the 6th, 22nd, and 32nd PCs) are presented in Figure \ref{fig_scree_img}. The noise contamination of the eigenvector and eigenimage becomes more significant as i increases. This contamination probably results in the convergence of $\delta v$ beyond the 22nd PC.

The PCs with $L$ smaller than 0.8~pc are excluded from the scaling relation if we adopt $N_\mathrm{sig}$ determined only by the scree plot method. Therefore, we cannot assess turbulence on small-scales. The uncertainty of $\alpha$ is also large since only four PCs are included in the fitting process (see Table \ref{tbl_PCA_scree}). In addition, our method covers the largest percentage of the total variation while avoiding noise contamination because the PCs beyond the 22nd PC are affected significantly by noise, as presented in Figure \ref{fig_scree_img}. These results imply that our method is more appropriate to constrain $N_\mathrm{sig}$, and thus, more precise properties of turbulence.

\begin{deluxetable}{lcccc}
\tablecolumns{5}
\tabletypesize{\scriptsize}
\tablecaption{The PCA results with different $N_\mathrm{sig}$\label{tbl_PCA_scree}}
\tablewidth{0pt}
\tablehead{
  \colhead{Included PCs}&\colhead{$N_\mathrm{sig}$}&\colhead{$p_\mathrm{var}$}&\colhead{$\log_{10}$($\delta v_0$)}&\colhead{$\alpha$}
}
\startdata
Cliff       	    	&   6&  63.2&   -0.12$\pm$0.04& 0.95$\pm$0.20\\
Orion A\tablenotemark{a}&   22& 72.1&   -0.09$\pm$0.02& 0.65$\pm$0.05\\
Cliff \& Scree 1	    &   32& 73.0&   -0.15$\pm$0.03& 0.49$\pm$0.06\\
\enddata
\tablenotetext{a}{The result for the PCs selected via our method. The values are the same as that presented in Table \ref{tbl_PCA_clouds}.}
\end{deluxetable}

\clearpage
\bibliographystyle{aasjournal.bst}
\bibliography{TIMES_II.bib}

\end{document}